\documentclass[aps,prl,10pt,twocolumn,footinbib,floatfix,showpacs,superscriptaddress]{revtex4-1}
\usepackage[colorlinks=true,citecolor=blue,linkcolor=blue,urlcolor=blue]{hyperref}
\usepackage{amsmath}
\usepackage{amssymb}
\usepackage{graphicx,amsmath,amsfonts,bm}
\usepackage{epstopdf}
\usepackage{bm}
\usepackage{color}
\usepackage{relsize}
\usepackage{dsfont}
\usepackage{braket}
\usepackage[caption=false]{subfig}
\usepackage{hyperref}
\usepackage[all]{hypcap}
\usepackage[T1]{fontenc}
\usepackage{dsfont}
\usepackage{mathbbol}
\usepackage{soul}

\renewcommand{\i}{\ensuremath{\mathrm{i}}}
\newcommand{\e}{\ensuremath{\mathrm{e}}}
\renewcommand{\d}{\ensuremath{\mathrm{d}}}

\begin{document}
\title{Testing Topological Protection of Edge States in Hexagonal Quantum Spin Hall Candidate Materials}

\author{Fernando Dominguez}\thanks{The two first authors contributed equally.}
\affiliation{Institute for Theoretical Physics and Astrophysics, TP4, University of W\"{u}rzburg, Am Hubland, 97074 W\"{u}rzburg, Germany}
\author{Benedikt Scharf}\thanks{The two first authors contributed equally.}
\affiliation{Institute for Theoretical Physics and Astrophysics, TP4, University of W\"{u}rzburg, Am Hubland, 97074 W\"{u}rzburg, Germany}
\author{Gang Li}
\affiliation{School of Physical Science and Technology, ShanghaiTech University, Shanghai 201210, China}
\author{J\"org Sch\"afer}
\affiliation{Physikalisches Institut and R\"ontgen Center for Complex Material Systems, University of W\"{u}rzburg, Am Hubland, 97074 W\"{u}rzburg, Germany}
\author{Ralph Claessen}
\affiliation{Physikalisches Institut and R\"ontgen Center for Complex Material Systems, University of W\"{u}rzburg, Am Hubland, 97074 W\"{u}rzburg, Germany}
\author{Werner Hanke}
\affiliation{Institute for Theoretical Physics and Astrophysics, TP1, University of W\"{u}rzburg, Am Hubland, 97074 W\"{u}rzburg, Germany}
\author{Ronny Thomale}
\affiliation{Institute for Theoretical Physics and Astrophysics, TP1, University of W\"{u}rzburg, Am Hubland, 97074 W\"{u}rzburg, Germany}
\author{Ewelina M. Hankiewicz}
\affiliation{Institute for Theoretical Physics and Astrophysics, TP4, University of W\"{u}rzburg, Am Hubland, 97074 W\"{u}rzburg, Germany}

\date{\today}

\begin{abstract}
We analyze the detailed structure of topological edge mode protection occurring in hexagonal quantum spin Hall (QSH) materials. We focus on bismuthene, antimonene, and arsenene on a SiC substrate, which, due to their large bulk gap, may offer new opportunities for room-temperature QSH applications. While time reversal symmetry is responsible for the principal symmetry protected character of QSH states, the hexagonal edge terminations yield further aspects of crystal symmetry which affect the topological protection. We show that armchair QSH edge states remain gapless under an in-plane magnetic field in the direction along the edge, a hallmark of their topological crystalline protection. In contrast, an out-of-plane magnetic field opens a gap of the order of a few meV within realistic ranges of the parameters. We use these intriguing signatures of armchair QSH edge states to predict experimentally testable fingerprints of their additional topological crystalline character and their helicity emerging in tunneling spectroscopy and ballistic magnetotransport.
\end{abstract}


\maketitle

\begin{figure}[t]
\centering
\includegraphics*[width=8cm]{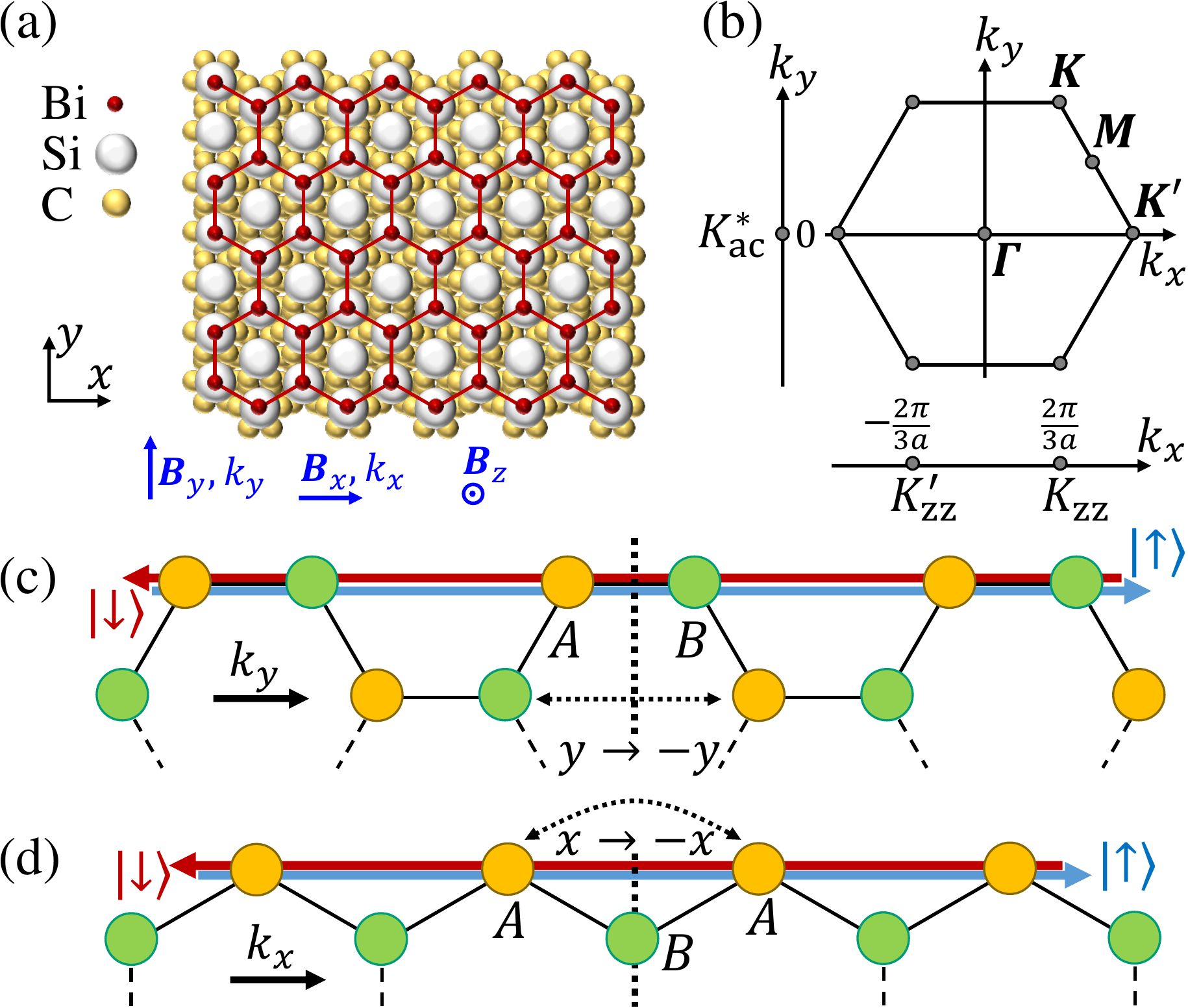}
\caption{(a) Schematic representation of a bismuthene layer on a SiC substrate. (b) Brillouin zone and nanoribbon $k$ spaces, denoted as $k_x$ and $k_y$ for ZZ and AC nanoribbons, respectively. The projections of the high-symmetry points to the nanoribbon $k$ spaces are shown in~(b). (c) AC and (d) ZZ edges and the effect of the reflection symmetries on the $A$ and $B$ sublattices.}\label{fig:Scheme}
\end{figure}

{\it Introduction} --- Dissipationless edge currents in quantum spin Hall (QSH) systems offer unique opportunities for novel device applications~\cite{Hasan2010:RMP,*Qi2011:RMP}. However, one of the main limiting factors of QSH materials is their small bulk band gap requiring cryogenic temperatures~\cite{Kane2005:PRL,*Kane2005:PRL2,Bernevig2006:S,Koenig2007:S,Knez2011:PRL}. In this context, a major step towards the realization of room-temperature QSH applications is bismuthene~\cite{Reis2017:S}, that is, Bi atoms arrayed in a honeycomb lattice, on a SiC(0001) substrate (Fig.~\ref{fig:Scheme}). Here, the SiC substrate stabilizes the two-dimensional (2D) layer of Bi atoms and shifts the $p_z$ orbitals of Bi away from the low-energy sector. As a consequence of this orbital filtering, the low-energy physics of the system is governed by the Bi $p_x$ and $p_y$ orbitals~\cite{Wu2008:PRL,Zhang2011:PRA,Zhou2018:NPJQM}, which in turn give rise to a large atomic on-site spin-orbit coupling (SOC). Such a mechanism, also predicted for Sb or As on a SiC substrate~\cite{Li2018:arxiv}, does not only allow for significantly larger bulk gaps ($\sim0.8\,$eV) compared to HgTe~\cite{Bernevig2006:S,Koenig2007:S,Koenig2008:JPSJ,Buettner2011:NP,Roth2009:S,Bruene2012:NP} and InAs/GaSb~\cite{Knez2011:PRL} quantum wells (QWs) or WTe$_2$~\cite{Fei2017:NP,Jia2017:PRB,Wu2018:S} layers, but also compared to other hexagonal layers predicted to exhibit helical states, such as jacutingaite~\cite{Marrazzo2018:PRL}, silicene~\cite{Liu2011:PRL,Vogt2012:PRL,Li2017:NL,Quhe2012:SR}, germanene~\cite{Liu2011:PRL,Zhang2016:PRL}, stanene~\cite{Xu2013:PRL,Ji2015:SR}, $[(\text{Bi}_{4}\text{Rh})_3\text{I}]^{2+}$~\cite{Rasche2013:NM,Pauly2015:NP}, or graphene on WS$_2$~\cite{Wang2015:NC,Gmitra2016:PRB,*Frank2018:PRL}.

In QSH systems, time-reversal symmetry (TRS) prevents helical edge states from mixing and gives rise to a protected crossing point in the edge spectrum and a quantized longitudinal conductance. If TRS is broken by an in-plane magnetic field, $z$-spin-polarized QSH states are expected to mix and a significant gap opens in the edge states ~\cite{Kane2005:PRL,*Kane2005:PRL2,Bernevig2006:S,Maciejko2010:PRB}. In contrast, a perpendicular magnetic field $B_z$ (see Fig.~\ref{fig:Scheme} for the coordinate axes) mixes opposite helicities only indirectly via Rashba or Dresselhaus SOC~\cite{Maciejko2010:PRB,Beugeling2012:PRB2}. Hence, in the presence of small Rashba and Dresselhaus SOC, nearly gapless helical edge states persist for finite $B_z$ in the ballistic limit. This occurs, for example, in QW-based QSH systems, such as symmetric HgTe QWs~\cite{Rothe2010:NJoP,Tkachov2010:PRL,*Tkachov2012:PhysicaE,Ilan2012:PRL,Chen2012:PRB,Scharf2012:PRB2,*Scharf2015:PRB2,Kharitonov2016:PRB}. 

In this manuscript, we investigate the hierarchy of topological protection in general hexagonal QSH systems with particle-hole symmetry (PHS). We find a generic topological crystalline protection arising from the interplay of PHS and reflection symmetry along the armchair (AC) edge [Fig.~\ref{fig:Scheme}(c,d)]. This topological protection manifests itself in gapless AC edge states for any direction of the magnetic field $\bm{B}$. In contrast, zigzag (ZZ) QSH edge states show a finite gap opening for any direction of $\bm{B}$~\cite{Kane2005:PRL,*Kane2005:PRL2,Lado2014:PRL,Rachel2014:PRB} due to breaking of the reflection symmetry responsible for this protection. Remarkably, even after breaking all symmetries of the system, nanoribbons with AC QSH edge states exhibit a suppressed gap, reminiscent of their topological crystalline protection. We apply these results to bismuthene, antimonene, and arsenene on SiC with well controlled AC edge termination~\footnote{In contrast to many other honeycomb systems, the use of a terraced SiC substrate allows for an efficient control of the edge termination of bismuthene, antimonene, and arsenene on SiC and a well defined AC edge~\cite{Bandoh2014:MSF}.}. This allows us to predict experimentally testable signatures of topological crystalline protection and QSH edge state helicity in these materials and provides an alternative to non-local resistance measurements~\cite{Roth2009:S,Bruene2012:NP} for the confirmation of their topological nature.

{\it Model} --- We use an $8\times8$ tight-binding (TB) Hamiltonian describing the low-energy physics of bismuthene, antimonene, and arsenene on SiC~\cite{Reis2017:S,Li2018:arxiv}. For practical purposes, we henceforth use bismuthene parameters~\cite{Reis2017:S}. This Hamiltonian is dominated by the Bi $p_x$ and $p_y$ orbitals, localized either on the $A$ or $B$ sites of the honeycomb lattice [Fig.~\ref{fig:Scheme}(a)] and carrying spin $s=\uparrow/\downarrow$, 
\begin{equation}\label{Eq:TotalHamiltonian}
H=\left(\begin{array}{cc}
 H_{\uparrow\uparrow} & H_{\uparrow\downarrow} \\
 H_{\downarrow\uparrow} & H_{\downarrow\downarrow} \\
 \end{array}\right)\:
\end{equation}
with the basis $\ket{p^A_{x\uparrow}}$, $\ket{p^A_{y\uparrow}}$, $\ket{p^B_{x\uparrow}}$, $\ket{p^B_{y\uparrow}}$, $\ket{p^A_{x\downarrow}}$, $\ket{p^A_{y\downarrow}}$, $\ket{p^B_{x\downarrow}}$, $\ket{p^B_{y\downarrow}}$. Here, the spin-diagonal blocks
\begin{equation}\label{Eq:HamiltonianDiagonalBlock}
H_{\uparrow\uparrow/\downarrow\downarrow}=\left(\begin{array}{cccc}
 0 & \mp\i\lambda_\mathsmaller{\mathrm{SOC}} & h_{xx}^{AB} & h_{xy}^{AB} \\
 \pm\i\lambda_\mathsmaller{\mathrm{SOC}} & 0 & h_{yx}^{AB} & h_{yy}^{AB} \\
 \left(h_{xx}^{AB}\right)^* & \left(h_{xy}^{AB}\right)^* & 0 & \mp\i\lambda_\mathsmaller{\mathrm{SOC}} \\
 \left(h_{yx}^{AB}\right)^* & \left(h_{yy}^{AB}\right)^* & \pm\i\lambda_\mathsmaller{\mathrm{SOC}} & 0 \\
 \end{array}\right)
\end{equation}
contain nearest-neighbor hopping terms (in reciprocal space) $h_{ij}^{AB}=h_{ij}^{AB}(\bm{k})$ between sublattices $A$ and $B$ parametrized by Slater-Koster integrals~\cite{SM2}. Crucially, Eq.~(\ref{Eq:HamiltonianDiagonalBlock}) also includes a large effective on-site SOC between the $p_x$ and $p_y$ orbitals, $\lambda_\mathsmaller{\mathrm{SOC}}=435\,$meV, responsible for a large bulk band gap at the $\bm{K}/\bm{K}'$ points.

\begin{figure}[t]
\centering
\includegraphics*[width=8.5cm]{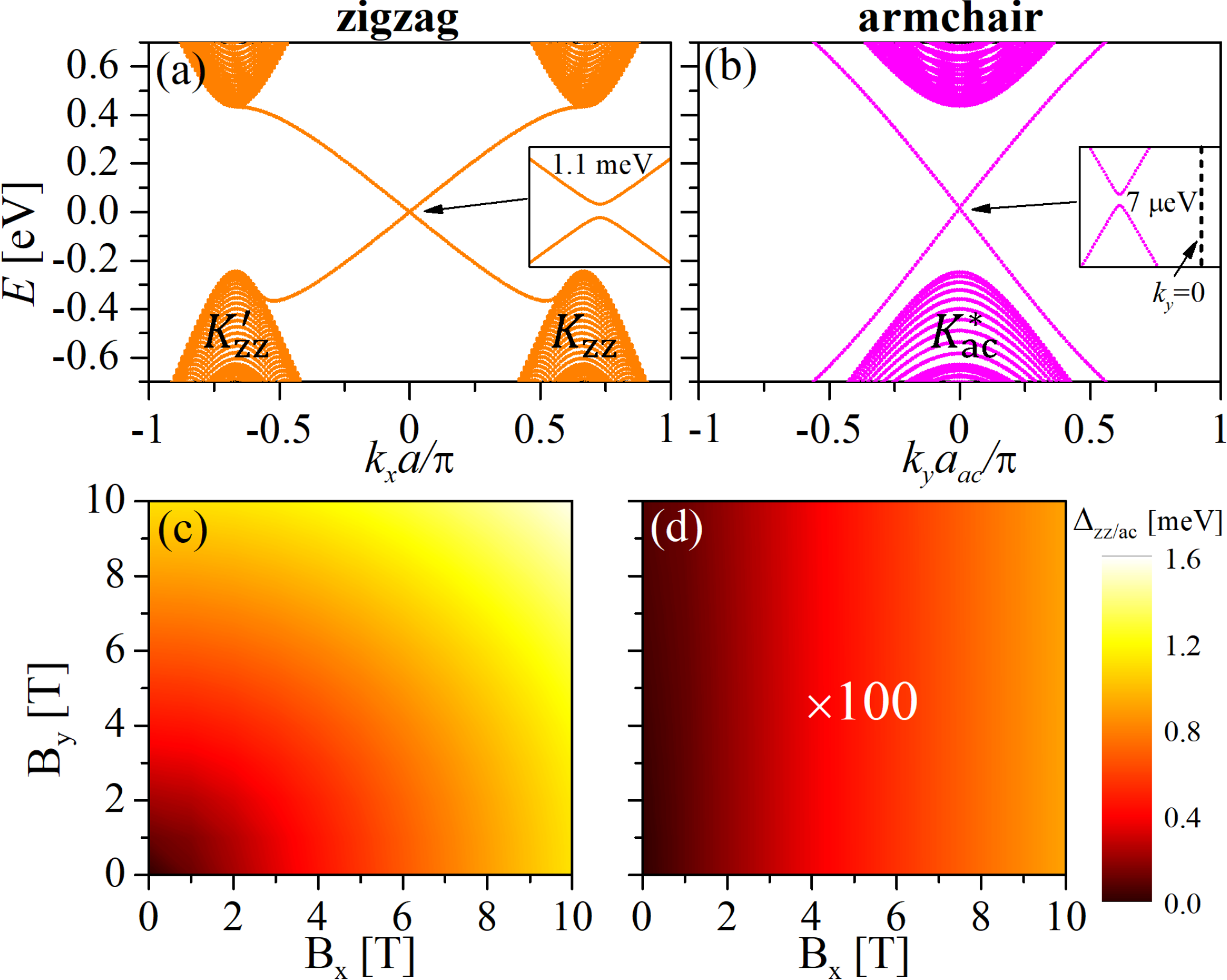}
\caption{Dispersions of (a) ZZ and (b) AC nanoribbons with a width of $N_{x/y}=100$ lattice sites for $B_x=10$ T. Insets in~(a) and~(b) show the gaps $\Delta_\text{zz}$ and $\Delta_\text{ac}$ opened between the QSH edge states at finite $B_x$. Panels~(c) and~(d) show the dependence of $\Delta_\text{zz}$ and $\Delta_\text{ac}$ on $B_x$ and $B_y$. Note that $\Delta_\text{ac}$ is multiplied by a factor 100. Here, $a_\mathrm{ac}=\sqrt{3}a$ is the length of an AC unit cell and $K^*_{ac}$ is shown in Fig.~\ref{fig:Scheme}(b).}\label{fig:InPlane}
\end{figure}

In addition, Rashba SOC enters in the off-diagonal terms $H_{\uparrow\downarrow}$, mixing both the spin and sublattice degrees of freedom. This term is proportional to the coupling constant $\lambda_\mathsmaller{\mathrm{R}}$ and lifts the degeneracy of the valence bands at $\bm{K}/\bm{K}'$ resulting in a valence band splitting of $12\lambda_\mathsmaller{\mathrm{R}}\approx 0.4\,$eV, also observed experimentally~\cite{Reis2017:S}. For the specific form of $H_{\uparrow\downarrow}$, we refer to Refs.~\cite{Reis2017:S,SM2}.

Magnetic fields induce orbital effects [Peierls phase in Eq.~(\ref{Eq:TotalHamiltonian})] and the Zeeman term
\begin{equation}\label{Eq:HamiltonianZeeman}
H_\mathrm{Z}=\mu_xB_x+\mu_yB_y+\mu_zB_z.
\end{equation}
We compute $H_\mathrm{Z}$ by applying L\"owdin perturbation theory~\cite{Winkler2003,Graf1995:PRB} around the $\bm{K}$ point to an ab-initio-based $52\times52$ Hamiltonian~\cite{Hohenberg1964:PR,Kohn1965:PR,Kresse1996:PRB,Bloechl1994:PRB,Perdew1996:PRL,Monkhorst1976:PRB,Marzari1997:PRB} and downfolding 
this Hamiltonian to the 8 bands of our TB Hamiltonian~(\ref{Eq:TotalHamiltonian}). In Eq.~(\ref{Eq:HamiltonianZeeman}), the magnetic moments $\mu_{x/y}=g_{\parallel}\mu_\mathsmaller{\mathrm{B}}\bm{1}\otimes s_{x/y}/2$ are $8\times 8$ matrices, $\mu_\mathsmaller{\mathrm{B}}$ is the Bohr magneton, $g_{\parallel}\approx2$, $s_i$ are spin Pauli matrices and $\bm{1}$ is the $4\times4$ unit matrix. $\mu_z$ is also $8\times 8$ matrix (see Ref.~\cite{SM2}), which has a non-diagonal spin structure because the higher-energy bands are spin-quantized along different axes than the 8 bands of our TB model.

{\it Testing the topological protection} --- We diagonalize the AC and ZZ nanoribbon Hamiltonians obtained from the corresponding discretization of the bulk Hamiltonian given in Eqs.~\eqref{Eq:TotalHamiltonian}-\eqref{Eq:HamiltonianZeeman}~\cite{SM2}. We observe (not shown) that in the absence of Rashba SOC, $\lambda_\text{R}=0$, the action of an in-plane magnetic field ($\bm{B}_\parallel$) always opens a gap in QSH ZZ edge states, while it never does for QSH AC edge states. In turn, for $\lambda_\text{R}\neq0$, a magnetic field along the AC edge does not open a gap, while a finite $\Delta_\text{ac}$, two orders of magnitude smaller than $\Delta_\text{zz}$ [Figs.~\ref{fig:InPlane}(a,b)], opens for the other direction of $\bm{B}_\parallel$. This strong dependence of $\Delta_\text{ac}$ on the direction of $\bm{B}_\parallel$ can be seen in Fig.~\ref{fig:InPlane}(d), very different than for ZZ edges [Fig.~\ref{fig:InPlane}(c)].

In order to understand these numerical observations, we study the symmetry class and the topological invariant of Eqs.~\eqref{Eq:TotalHamiltonian}-\eqref{Eq:HamiltonianZeeman} with $\lambda_\mathsmaller{\mathrm{R}}=0$ and $\bm{B}_\parallel\neq 0$. The presence of $\bm{B}_\parallel$ breaks TRS, leaving only PHS. Then, $H$ belongs to symmetry class D. In addition, crystal symmetries, including reflection, rotation, etc can modify and/or extend the tenfold classification of topological insulators, leading to so-called topological crystalline insulators~\cite{Fu2011a,Hsieh2012a,Chiu2013a,Shiozaki2014a,Ando2015a,Chiu2016a}, observed by several groups~\cite{Xu2012a,Tanaka2012a,Dziawa2012a,Sessi2016a}. Here, the bulk Hamiltonian exhibits two reflection symmetries $\mathcal{R}(x)$ and $\mathcal{R}(y)$, acting on the Hamiltonian as
\begin{align}
\mathcal{R}(i)H(\overline{\bm{k}})\mathcal{R}^{-1}(i)=H(\bm{k}),
\end{align}
where $i=x,y$, $\bm{k}=(k_x,k_y)$ and $\overline{\bm{k}}$ is equal to $\bm{k}$ except for its $i$th component, which is reflected ($k_i\rightarrow -k_i$).
The key difference between $\mathcal{R}(x)$ and $\mathcal{R}(y)$ is that $\mathcal{R}(y)\propto \sigma_x$ mixes $A$ and $B$ sublattices ($\sigma_i$ are Pauli matrices in sublattice space), while $\mathcal{R}(x)\propto \sigma_0$ is diagonal in this subspace. Following a standard procedure, we find that only the combination of PHS and $\mathcal{R}(y)$ leads to a non-trivial mirror topological invariant, the mirror Chern number $MZ_2$~\cite{Teo2008a,Chiu2013a,Chiu2014:PRB,SM2}.
\emph{However, why are ZZ nanoribbons showing an opening of a gap?} At this point, it is important to realize that not all boundary conditions are compatible with $\mathcal{R}(y)$. Indeed, ZZ boundary conditions ($\propto \sigma_z$) do not preserve $\mathcal{R}(y)\propto \sigma_x$ and thus, $\bm{B}_\parallel$ can open a gap. In turn, $\mathcal{R}(y)$ is compatible with AC boundary conditions and therefore, the resulting crossing is topologically protected against $\bm{B}_\parallel$~\cite{SM2}.

\begin{figure}[t]
\centering
\includegraphics*[width=8.5cm]{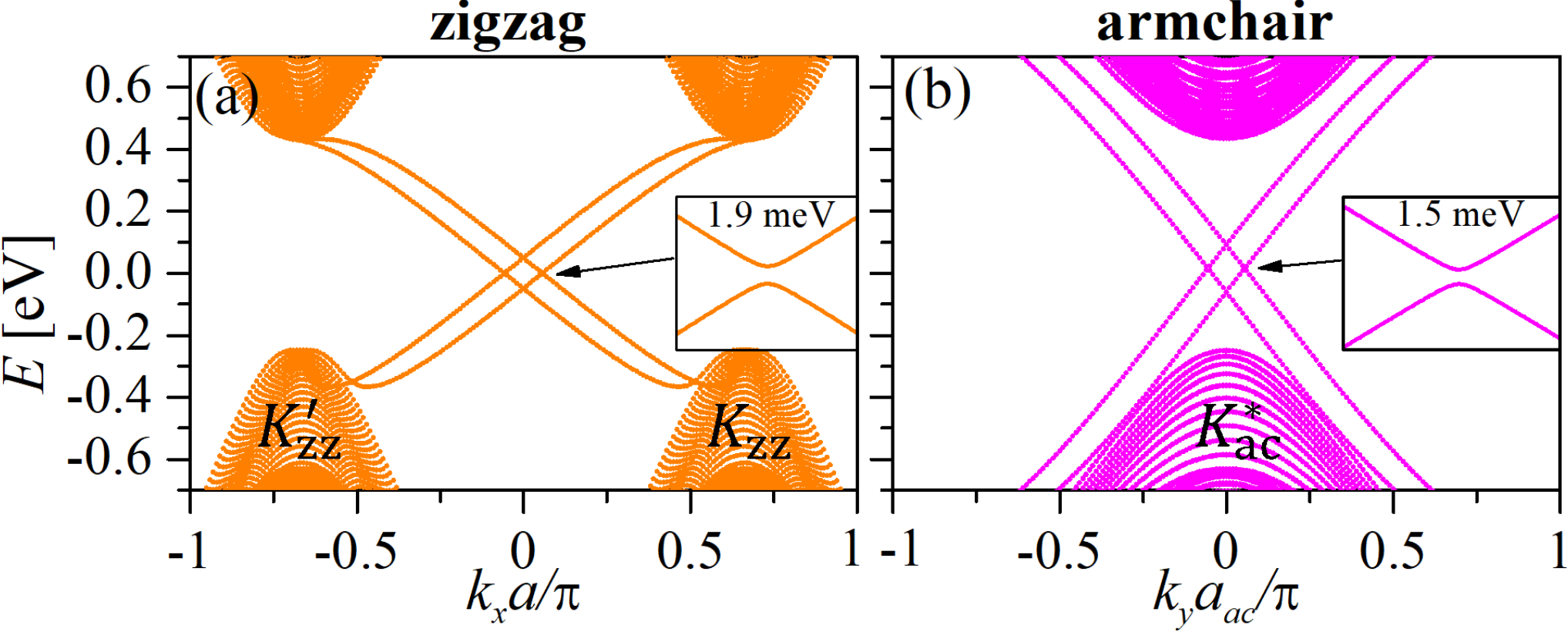}
\caption{Dispersions of (a) ZZ and (b) AC nanoribbons with $N_{x/y}=100$ lattice sites for $B_z=10$ T. The insets in~(a) and~(b) show the gaps $\Delta_\text{zz}$ and $\Delta_\text{ac}$ opened between the QSH states at finite $B_z$.}\label{fig:OutOfPlane}
\end{figure}

The crystalline topological protection discussed so far can be extended to all hexagonal QSH materials exhibiting PHS, such as the Kane-Mele Hamiltonian. Note, however, that PHS is present only approximately in practice. In the presence of terms breaking PHS, we expect $\bm{B}_\parallel$ to open a gap. In bismuthene, the main contribution breaking PHS is Rashba SOC. Thus, we now estimate $\Delta_\text{ac}$ opened by $\bm{B}_\parallel$, when $\lambda_\mathsmaller{\mathrm{R}}\neq 0$. To do so, we expand Eq.~\eqref{Eq:TotalHamiltonian} around $\bm{K}/\bm{K}'$~\cite{SM2}, yielding $H_\mathrm{eff}= H_0+H_\text{R}$, with the spin-diagonal contribution
\begin{align}
H_0=\hbar v_\text{F}(q_x\sigma_x\tau_z+q_y\sigma_y)+\lambda_\text{SOC}\sigma_zs_z\tau_z,
\label{Eq:EffHamiltonian}
\end{align}
and the non-diagonal contribution due to Rashba SOC
\begin{align}\label{Eq:EffHamiltonianR}
H_\text{R}= 3\lambda_\mathsmaller{\mathrm{R}}(&\sigma_x s_y \tau_z-\sigma_y s_x)\nonumber\\
& + \sqrt{3} q_y a \lambda_\mathsmaller{\mathrm{R}}(\sigma_x s_x +\sigma_y s_y\tau_z),
\end{align}
where $q_x$ and $q_y$ are momenta measured from $\bm{K}/\bm{K}'$, $v_\text{F}$ is the Fermi velocity, and $\sigma_i$, $s_i$, and $\tau_i$ are Pauli matrices for sublattice, spin, and valley, respectively. Here, the basis is given by $\{|\tau\i p_x^B+p_y^B\rangle,|\tau\i p_x^\text{A}+p_y^\text{A}\rangle\}$, with $\tau=\pm1$ corresponding to the $\bm{K}/\bm{K'}$ valleys.

\begin{figure}[t]
\centering
\includegraphics*[width=8.5cm]{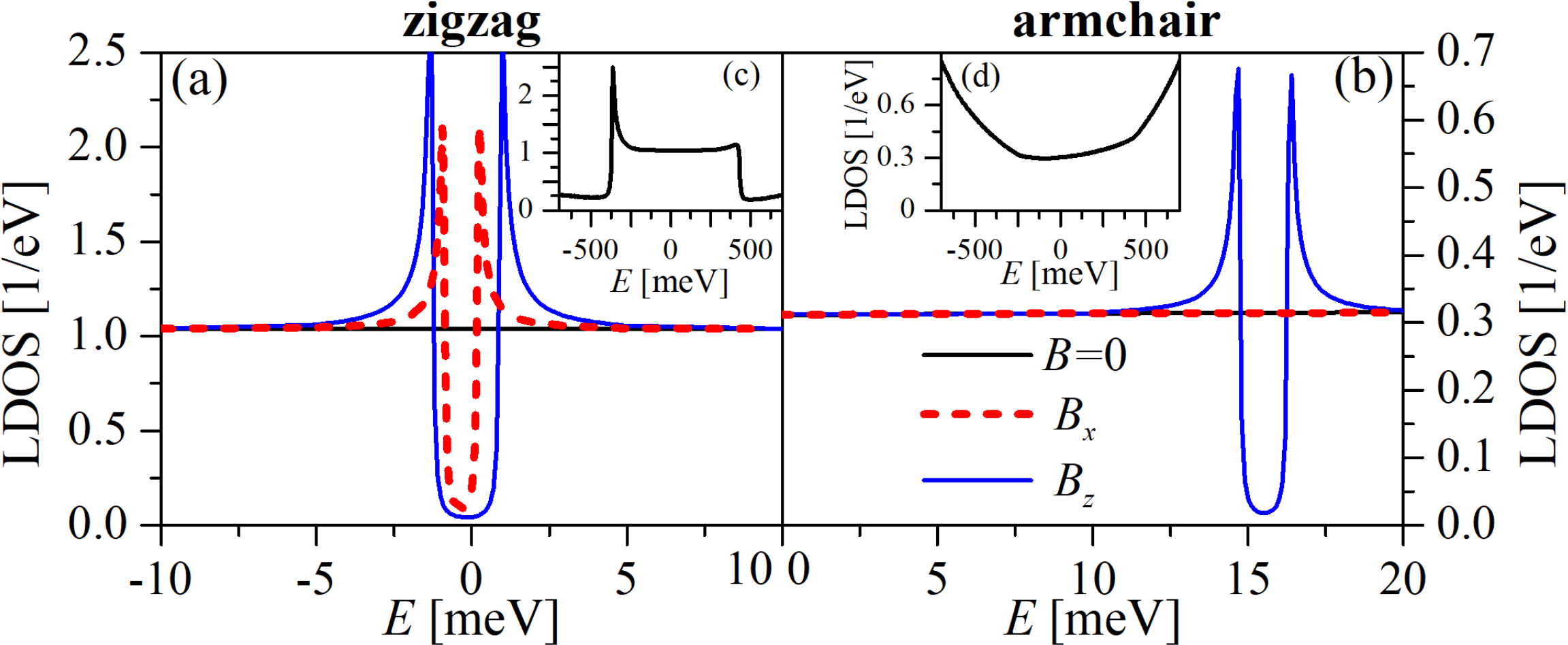}
\caption{Edge LDOS of (a) ZZ and (b) AC nanoribbons with $N_{x/y}=100$ lattice sites close to $E=0$ for zero and finite $\bm{B}$. Insets~(c) and~(d) show the edge LDOS at $B=0$ in a wider energy window. We use a Lorentzian broadening $\Gamma=50\,\mu$eV in (a,b) and $\Gamma=5\,$meV in (c,d).}\label{fig:LDOS}
\end{figure}

Analytical results determined from $H_\mathrm{eff}$ with AC boundary conditions~\cite{Brey2006a} show no gap opening due to $\mu_y B_y$ because this direction is (trivially) protected by reflection symmetry $\mathcal{R}(y)=\sigma_x s_y$, which takes the role of the helicity operator here. In turn, a Zeeman term in $x$-direction opens a second-order gap in $\lambda_\text{R}$, scaling as~\cite{SM2}
\begin{align}
\Delta_\text{ac}\approx \frac{3\sqrt{3}a \lambda_\text{R}^2}{\lambda_\text{SOC} \hbar v_\text{F}} \mu_\mathsmaller{\mathrm{B}}B_x \sim 10^{-2}\mu_\mathsmaller{\mathrm{B}}B_x.
\end{align}
Both responses to $\mu_x B_x$ and $\mu_y B_y$ are in good agreement with numerical observations and explain the results shown in Fig.~\ref{fig:InPlane}(d), where, for example, $B_x=10\,$T ($\mu_\mathsmaller{\mathrm{B}}B_x\sim 0.6\,$meV) yields $\Delta_\text{ac}\approx7\,\mu$eV. Remarkably, even after breaking all symmetries, AC QSH edge states show gaps two orders smaller than those at ZZ edges for in-plane fields.

Following a similar reasoning, $\mu_z B_z$ together with $H_\text{R}$ can also open a gap since both contributions break PHS. Here, $\mu_z B_z$ dominates the gap opening due to its non-diagonal structure. Thus, we find comparable AC and ZZ gaps of around a few meV for $B_z=10\,$T ($\Delta_\text{ac}\sim\Delta_\text{zz}$, see Fig.~\ref{fig:OutOfPlane}). Both, $\Delta_\text{ac}$ and $\Delta_\text{zz}$ exhibit a $B_z$-linear dependence with slopes of around $0.2\,$meV/T. This is the usual behavior expected also for QSH insulators with strong Rashba SOC like InAs/GaSb QWs~\cite{Knez2011:PRL}. 
Next, we study signatures of $\Delta_\text{ac}$ and $\Delta_\text{zz}$ and the topological nature of the QSH states in experimentally accessible quantities.

{\it How to detect the topological nature of the edge states?} --- In Fig.~\ref{fig:LDOS}, we show the edge local density of states (LDOS) computed from the TB model as a function of energy for ZZ and AC nanoribbons. At $B=0$, the linear dispersion of the edge states gives rise to a flat LDOS. At finite $\bm{B}$, a dip corresponding to $\Delta_\text{ac}$ or $\Delta_\text{zz}$ arises in the edge LDOS, while away from this gap, the edge LDOS is not significantly altered even for $B=10\,$T. Depending on the broadening limited by the parameters of the experimental setup, such as temperature $T$, $\Delta_\text{ac}$ and $\Delta_\text{zz}$ can be resolved in LDOS measurements. With $\Delta_\text{ac}$ and $\Delta_\text{zz}$ of a few meV, predicted at $B=10\,$T for ZZ (any $\bm{B}$) and AC ribbons ($B_z$), we expect that such gaps could be measured experimentally. This is illustrated in Figs.~\ref{fig:LDOS}(a,b), where we have chosen a broadening $\Gamma=50\,\mu$eV. Note that the opening of the gap in Fig.~\ref{fig:LDOS}(b) occurs away from $E=0$ due to Rashba SOC breaking PHS. 
Remarkably, for AC QSH edge states we predict a tiny non-measurable gap of $7\,\mu$eV ($\approx80\,$mK) with $B_x$ and no gap for $B_y$ and consequently a constant signal in the LDOS for both cases. Employing scanning tunneling spectroscopy~\cite{Reis2017:S} to monitor the LDOS for different $\bm{B}$ orientations, the distinct behavior of AC QSH edge states could serve as a smoking gun to distinguish these states from trivial Rashba edge states, such as those observed in bismuth thin films~\cite{Takayama2015:PRL}: For the latter, one would expect a dip in the LDOS as a function of $E$, independent of the orientation of $\bm{B}$, whereas the LDOS of AC QSH edge states exhibits different responses to in-plane $\bm{B}$ (flat LDOS) and out-of-plane $\bm{B}$ (dip in LDOS).

If we focus on a larger energy window, the ZZ edge LDOS also exhibits a clear asymmetry arising from the Rashba-split valence bands. In addition, we can observe signatures of the reduced (1D) dimensionality of the edge states: The van Hove singularities result from bending the linear-dispersion edge state into the continuum as shown in Fig.~\ref{fig:LDOS}(c) [compare Figs.~\ref{fig:InPlane}(a) and~\ref{fig:OutOfPlane}(a)]. Moreover, we observe qualitative differences comparing the LDOS for AC and ZZ edges. While the AC edge LDOS increases for energies outside the bulk band gap [Fig.~\ref{fig:LDOS}(d)], the ZZ edge LDOS drops significantly at these energies [Fig.~\ref{fig:LDOS}(c)]. This behavior reflects the fact that AC edge states do not immediately merge with the bulk states at the band edges [compare Figs.~\ref{fig:InPlane}(b) and~\ref{fig:OutOfPlane}(b)]. Hence, measurements of the edge LDOS can also elucidate the nature of the boundaries, even at $B=0$. Indeed, our results for the AC edge LDOS in Fig.\ref{fig:LDOS}(d) are qualitatively similar to the LDOS observed experimentally~\cite{Reis2017:S}.

\begin{figure}[t]
\centering
\includegraphics*[width=8.5cm]{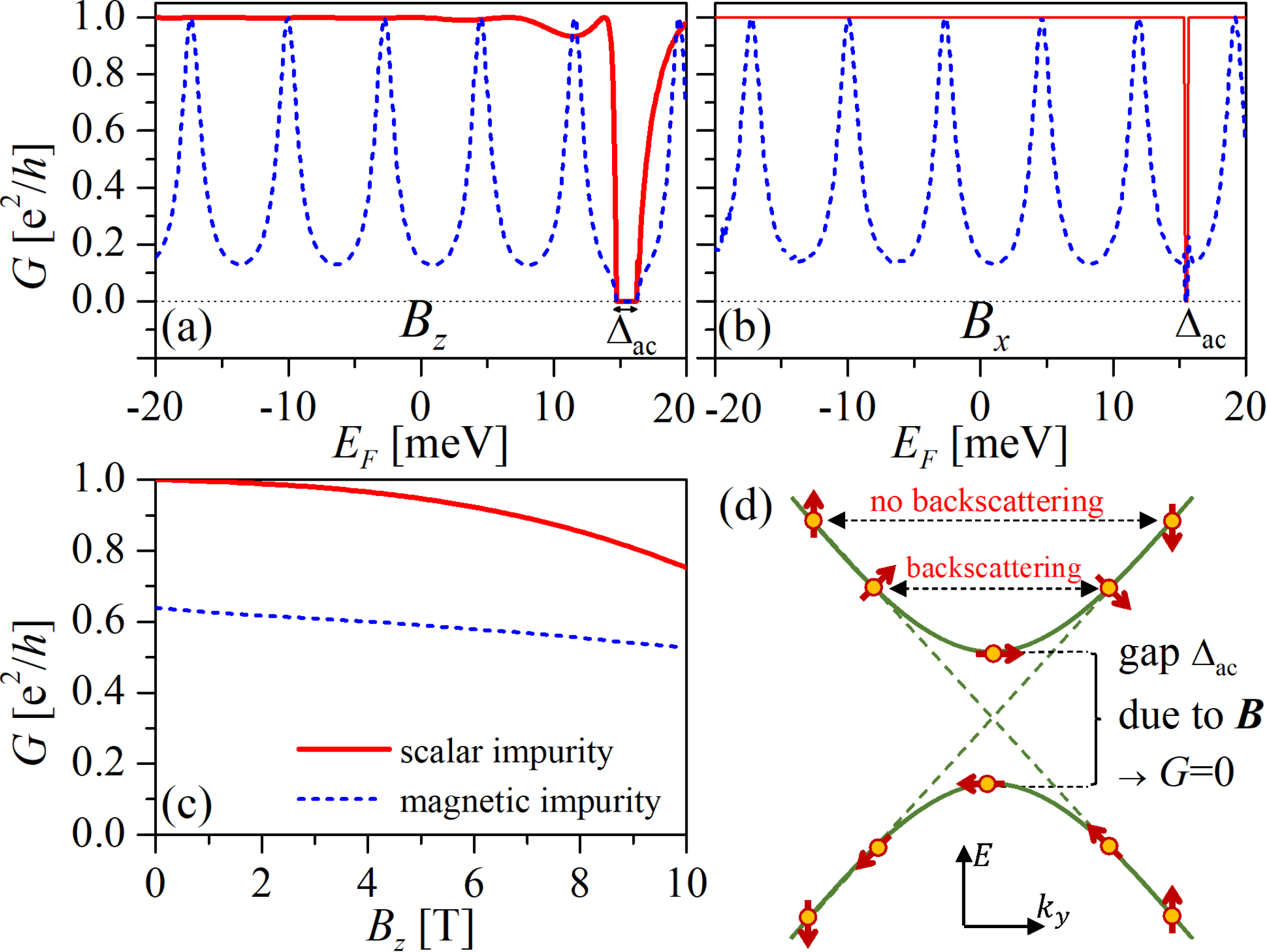}
\caption{Conductance $G$ of a single AC edge as a function of the Fermi energy $E_F$ for (a) $B_z=10\,$T and (b) $B_x=10\,$T with two scalar or magnetic impurities present at the edge. (c) Dependence of $G$ on $B_z$ for $E_F=17\,$meV and two impurities. In all panels, the impurities are separated by $d_\mathrm{imp}=100$ nm from each other and have a strength of $V_\mathrm{imp}=10\,${eV\AA} each in (a,b) and $V_\mathrm{imp}=5\,${eV\AA} each in (c). (d) Schematic edge-state dispersion with $\Delta_\text{ac}$ opened due to finite $\bm{B}$ [see~(a,b)] and schematic spin expectation values $\braket{\bm{s}}$. Note that the actual values and directions of $\braket{\bm{s}}$ depend on the strengths of $\bm{B}$ and Rashba SOC.}\label{fig:Conductance}
\end{figure}

Another quantity exhibiting signatures of $\Delta_\text{ac}$ or $\Delta_\text{zz}$ and the spin polarization/helicity of the QSH edge states is the magnetoconductance $G$ in the ballistic regime, shown in Fig.~\ref{fig:Conductance}. Here, we compute $G$ of a single AC edge at finite $\bm{B}$ via the Fisher-Lee relation~\cite{Fisher1981:PRB,SM2}. In the absence of impurities, the conductance of a single AC edge is perfectly quantized at $e^2/h$ inside the bulk gap for $B=0$.
This quantization remains for finite $\bm{B}$, even if scalar impurities are included~\cite{Tkachov2012:PRB}. The conductance deviates from its quantized value only for the energy window corresponding to the $\bm{B}$-induced $\Delta_\text{ac}$, where eventually $G$ drops to zero [Figs.~\ref{fig:Conductance}(a,b)]. Here, no propagating states are available as illustrated in Fig.~\ref{fig:Conductance}(d). If both $\bm{B}$ and scalar impurities are present, we observe a small reduction with respect to $e^2/h$ close to the gap opening [Fig.~\ref{fig:Conductance}(a)]. This small deviation reflects the fact that the counter-propagating states are not perfectly spin-polarized and can thus scatter even at scalar impurities for a Fermi energy $E_F$ around $\Delta_\text{ac}$ at finite $\bm{B}$. The suppression of $G$ close to the gap is also shown in Fig.~\ref{fig:Conductance}(c), which monitors the $B_z$-dependence of $G$ at fixed $E_F$.

The presence of magnetic impurities significantly reduces $G$, for both $B=0$ and finite $\bm{B}$, independent of the $\bm{B}$ direction [dashed blue lines in Figs.~\ref{fig:Conductance}(a-c)]. Here, $G$ decays exponentially with the number of magnetic impurities~\cite{SM2}. Similar to HgTe QWs~\cite{Tkachov2010:PRL}, $G$ exhibits Fabry-Perot-type oscillations if multiple magnetic impurities are situated at the edge [Figs.~\ref{fig:Conductance}(a,b)]. The behavior of $G$ due to impurities, which could be measured in a two-terminal setup, allows us to distinguish between QSH and trivial edge states, since for the latter $G$ would not be quantized, even at $B=0$.

{\it Outlook} --- We study the hierarchy of topological protection in general hexagonal QSH systems with PHS. We find a generic topological crystalline protection arising from the interplay of PHS and reflection symmetry along AC edges. This topological protection manifests itself in gapless AC edge states for any direction of $\bm{B}$. In contrast, ZZ boundary conditions break the reflection symmetry responsible for this protection and give rise to more usual QSH edge states with a finite gap opening for any direction of $\bm{B}$. 
Remarkably, even after breaking reflection and PHS, AC QSH states show a suppressed gap, reminiscent of crystalline protection. This special behavior of QSH edge states in $\bm{B}$ opens the possibility for testing crystalline topological protection and spin helicity in new candidates for QSH systems such as bismuthene, antimonene, and arsenene on SiC. Further, our results have potential applications in spintronics to manipulate/switch spin currents and hybrid superconductor/honeycomb QSH systems, where manipulating the edge-state gap has important consequences for the control of Majorana modes in Josephson junctions~\cite{Kuzmanovski2016a}.

\begin{acknowledgments}
{\it Acknowledgments} --- We thank Fernando de Juan, Tobias Frank, Dimitri Jungblut, Felix Reis, Grigory Tkachov, and Bj\"orn Trauzettel for valuable discussions. This work was supported by the German Science Foundation (DFG) via Grant No. SFB 1170 ``ToCoTronics'', by the ENB Graduate School on Topological Insulators, and by the European Research Council via Grant No. ERC-StG-Thomale-TOPOLECTRICS-336012.
\end{acknowledgments}

\bibliography{BibTopInsAndTopSup}

\section{Tight-binding model for nanoribbons}\label{Sec:TBnr}
For brevity, the tight-binding (TB) Hamiltonian $H=H(\bm{k})$ given by Eqs.~(1)-(3) is presented in reciprocal space in the main text and can also be found in Ref.~\cite{Reis2017:S}. Using the basis ordering $\ket{p^A_{x\uparrow}}$, $\ket{p^A_{y\uparrow}}$, $\ket{p^B_{x\uparrow}}$, $\ket{p^B_{y\uparrow}}$, $\ket{p^A_{x\downarrow}}$, $\ket{p^A_{y\downarrow}}$, $\ket{p^B_{x\downarrow}}$, $\ket{p^B_{y\downarrow}}$ and choosing the direct lattice vectors as $\mathbf{a}_1=a\mathbf{e}_x$ and $\mathbf{a}_2=-(a/2)\mathbf{e}_x+(\sqrt{3}a/2)\mathbf{e}_y$ with the lattice constant $a=5.35$ \AA, the $8\times8$ TB Hamiltonian reads
\begin{equation}\label{Eq:TotalHamiltonian}
H=\left(\begin{array}{cc}
   H_{\uparrow\uparrow} & H_{\uparrow\downarrow} \\
   H_{\downarrow\uparrow} & H_{\downarrow\downarrow} \\
  \end{array}\right).
\end{equation}
The spin-diagonal blocks
\begin{equation}\label{Eq:HamiltonianDiagonalBlock}
H_{\uparrow\uparrow/\downarrow\downarrow}=\left(\begin{array}{cccc}
                              0 & \mp\i\lambda_\mathsmaller{\mathrm{SOC}} & h_{xx}^{AB} & h_{xy}^{AB} \\
                              \pm\i\lambda_\mathsmaller{\mathrm{SOC}} & 0 & h_{yx}^{AB} & h_{yy}^{AB} \\
                              \left(h_{xx}^{AB}\right)^* & \left(h_{yx}^{AB}\right)^* & 0 & \mp\i\lambda_\mathsmaller{\mathrm{SOC}} \\
                              \left(h_{xy}^{AB}\right)^* & \left(h_{yy}^{AB}\right)^* & \pm\i\lambda_\mathsmaller{\mathrm{SOC}} & 0 \\
                              \end{array}\right),\:
\end{equation}
\begin{equation}\label{Eq:HamiltonianDiagonalBlockSlaterKoster}
\begin{array}{l}
h_{xx}^{AB}=V^1_\mathsmaller{pp\pi}+\frac{3V^1_\mathsmaller{pp\sigma}+V^1_\mathsmaller{pp\pi}}{2}\,\e^{\i\frac{\sqrt{3}k_ya}{2}}\cos\left(\frac{k_xa}{2}\right),\\
h_{xy}^{AB}=h_{yx}^{AB}=\frac{\sqrt{3}\i\left(V^1_\mathsmaller{pp\sigma}-V^1_\mathsmaller{pp\pi}\right)}{2}\,\e^{\i\frac{\sqrt{3}k_ya}{2}}\sin\left(\frac{k_xa}{2}\right),\\
h_{yy}^{AB}=V^1_\mathsmaller{pp\sigma}+\frac{V^1_\mathsmaller{pp\sigma}+3V^1_\mathsmaller{pp\pi}}{2}\,\e^{\i\frac{\sqrt{3}k_ya}{2}}\cos\left(\frac{k_xa}{2}\right)
\end{array}
\end{equation}
contain nearest-neighbor hopping terms between sublattices $A$ and $B$ parametrized by the Slater-Koster integrals $V^1_\mathsmaller{pp\sigma}=2\,$eV and $V^1_\mathsmaller{pp\pi}=-210\,$meV. Crucially, Eq.~(\ref{Eq:HamiltonianDiagonalBlock}) also includes a large effective on-site SOC between the $p_x$ and $p_y$ orbitals, $\lambda_\mathsmaller{\mathrm{SOC}}=435\,$meV, giving rise to a band gap of $2\lambda_\mathsmaller{\mathrm{SOC}}$ between the conduction and valence bands at the $\bm{K}/\bm{K}'$ points (if no Rashba SOC is taken into account).

The Rashba-like SOC is given by
\begin{equation}\label{Eq:HamiltonianRashba}
H_{\uparrow\downarrow}=\left(H_{\downarrow\uparrow}\right)^\dagger=\lambda_\mathsmaller{\mathrm{R}}\left(\begin{array}{cccc}
                              0 & 0 & m_1 & m_2 \\
                              0 & 0 & m_2 & m_3 \\
                              m_4 & m_5 & 0 & 0 \\
                              m_5 & m_6 & 0 & 0 \\
                              \end{array}\right),\:
\end{equation}
\begin{equation}\label{Eq:HamiltonianRashbaParameters}
\begin{array}{l}
m_{1/4}=-2\sqrt{3}\i\,\e^{\pm\i\frac{\sqrt{3}k_ya}{2}}\sin\left(\frac{k_xa}{2}\right),\\
m_{2/5}=\pm\left\{1-\e^{\pm\i\frac{\sqrt{3}k_ya}{2}}\left[\cos\left(\frac{k_xa}{2}\right)\pm\sqrt{3}\sin\left(\frac{k_xa}{2}\right)\right]\right\},\\
m_{3/6}=\mp2\i\left[1-\e^{\pm\i\frac{\sqrt{3}k_ya}{2}}\cos\left(\frac{k_xa}{2}\right)\right]
\end{array}
\end{equation}
and lifts the degeneracy of the valence bands at the $\bm{K}/\bm{K}'$ points, resulting in a valence band splitting of $12\lambda_\mathsmaller{\mathrm{R}}$ with $\lambda_\mathsmaller{\mathrm{R}}=32\,$meV.

\begin{figure}[tb]
\centering
\includegraphics*[width=8.5cm]{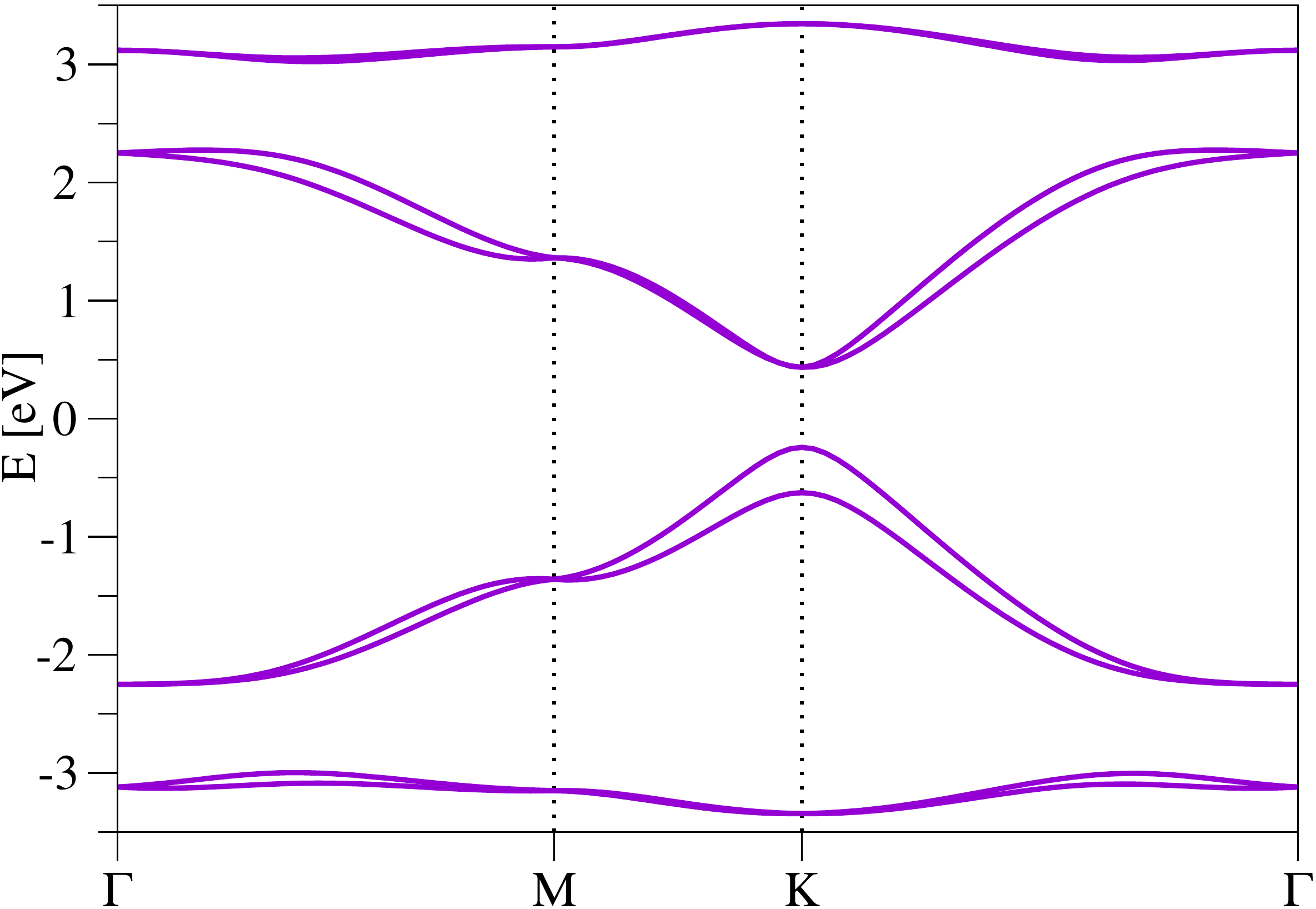}
\caption{Band structure of bismuthene on SiC as calculated using the tight-binding model given by Eqs.~(\ref{Eq:TotalHamiltonian})-(\ref{Eq:HamiltonianRashbaParameters}).}\label{fig:TBmodel}
\end{figure}

Equations~(\ref{Eq:TotalHamiltonian})-(\ref{Eq:HamiltonianRashbaParameters}) describe the band structure of bismuthene/SiC in an infinite two-dimensional (2D) plane at zero magnetic field $\bm{B}$ (shown in Fig.~\ref{fig:TBmodel}). Effective TB Hamiltonians for nanoribbons can then be obtained by one-dimensional (1D) Fourier transformations of $H(\bm{k})$ and $H_\mathrm{Z}$.

\subsection{Zigzag nanoribbons}
\begin{figure}[t]
\centering
\includegraphics*[width=8.5cm]{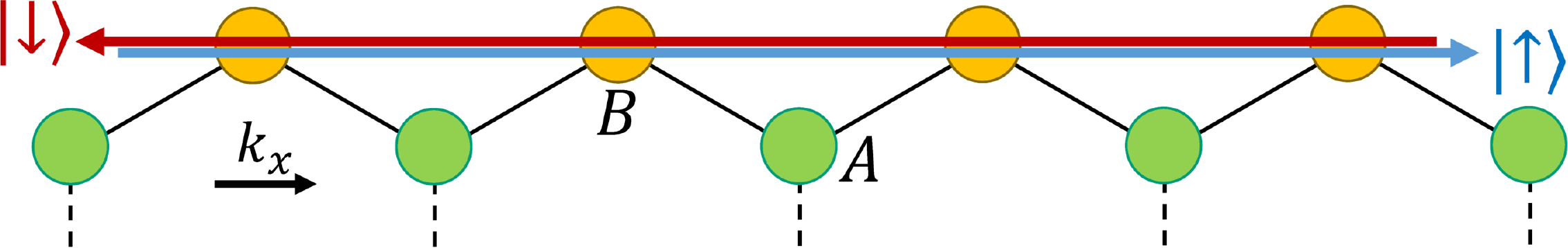}
\caption{Zigzag nanoribbons with QSH edge states. Symbols $\uparrow/\downarrow$ refer to states which have predominantly spins $\uparrow/\downarrow$.}\label{fig:ZZ}
\end{figure}

In particular, for our choice of basis vectors $\mathbf{a}_{1/2}$, the Fourier transform of $H(\bm{k})$ and $H_\mathrm{Z}$ with respect to $\sqrt{3}k_y/2$ yields nanoribbons with zigzag (ZZ) edges along the $x$-direction (see Fig.~\ref{fig:ZZ}), a finite number of lattice sites $n=1,..,N_y$ along the $y$-direction, and a good momentum quantum number $k_x$. Introducing the field operators 
\begin{equation}\label{Eq:TBzzFO}
\hat{\Psi}(k_x,n)=\left(\begin{array}{c}\hat{c}_{p^A_{x\uparrow}}(k_x,n)\\
\hat{c}_{p^A_{y\uparrow}}(k_x,n)\\
\hat{c}_{p^B_{x\uparrow}}(k_x,n)\\
\hat{c}_{p^B_{y\uparrow}}(k_x,n)\\
\hat{c}_{p^A_{x\downarrow}}(k_x,n)\\
\hat{c}_{p^A_{y\downarrow}}(k_x,n)\\
\hat{c}_{p^B_{x\downarrow}}(k_x,n)\\
\hat{c}_{p^B_{y\downarrow}}(k_x,n)\end{array}\right),
\end{equation}
which consist of the operators annihilating an electron at site $n$ in $y$-direction on sublattice $A/B$ with orbital $p_x/p_y$, spin $\uparrow/\downarrow$ and longitudinal momentum $k_x$, the Hamiltonian for ZZ nanoribbons can be written as
\begin{equation}\label{Eq:FullTBzz}
\hat{H}_\mathrm{zz}=\sum\limits_{k_x,n,n'}\hat{\Psi}^\dagger(k_x,n)H^\mathrm{zz}_{nn'}(k_x)\hat{\Psi}(k_x,n').
\end{equation}
Here, the matrix
\begin{equation}\label{Eq:TBzzMat}
H^\mathrm{zz}_{nn'}(k_x)=\left(\begin{array}{cc}
                           H^{\uparrow\uparrow}_{nn'}(k_x) & H^{\uparrow\downarrow}_{nn'}(k_x) \\
                           H^{\downarrow\uparrow}_{nn'}(k_x) & H^{\downarrow\downarrow}_{nn'}(k_x) \\
                           \end{array}\right)
\end{equation}
is given by
\begin{equation}\label{Eq:TBzzMatS}
\begin{array}{l}
H^{\uparrow\uparrow/\downarrow\downarrow}_{nn'}(k_x)=\\
\quad\\
                           \left(\begin{array}{cccc}
                              0 & \mp\i\lambda_\mathsmaller{\mathrm{SOC}}\delta_{n,n'} & h^+_{xx} & h^+_{xy} \\
                              \pm\i\lambda_\mathsmaller{\mathrm{SOC}}\delta_{n,n'} & 0 & h^+_{xy} & h^+_{yy} \\
                              h^-_{xx} & h^-_{xy} & 0 & \mp\i\lambda_\mathsmaller{\mathrm{SOC}}\delta_{n,n'} \\
                              h^-_{xy} & h^-_{yy} & \pm\i\lambda_\mathsmaller{\mathrm{SOC}}\delta_{n,n'} & 0
                           \end{array}\right)
\end{array}
\end{equation}
with
\begin{equation}\label{Eq:TBzzMatSe}
\begin{array}{l}
h^\pm_{xx}=V^1_\mathsmaller{pp\pi}\delta_{n,n'}+\frac{3V^1_\mathsmaller{pp\sigma}+V^1_\mathsmaller{pp\pi}}{2}\,\cos\left(\frac{k_xa}{2}\right)\delta_{n\pm1,n'},\\
h^\pm_{xy}=h^\pm_{yx}=\pm\frac{\sqrt{3}\i\left(V^1_\mathsmaller{pp\sigma}-V^1_\mathsmaller{pp\pi}\right)}{2}\,\sin\left(\frac{k_xa}{2}\right)\delta_{n\pm1,n'},\\
h^\pm_{yy}=V^1_\mathsmaller{pp\sigma}\delta_{n,n'}+\frac{V^1_\mathsmaller{pp\sigma}+3V^1_\mathsmaller{pp\pi}}{2}\,\cos\left(\frac{k_xa}{2}\right)\delta_{n\pm1,n'}
\end{array}
\end{equation}
and
\begin{equation}\label{Eq:TBzzMatR1}
\begin{array}{l}
H^{\uparrow\downarrow}_{nn'}(k_x)=\lambda_\mathsmaller{\mathrm{R}}\left(\begin{array}{cccc}
                              0 & 0 & m_1 & m_2 \\
                              0 & 0 & m_2 & m_3 \\
                              m_4 & m_5 & 0 & 0 \\
                              m_5 & m_6 & 0 & 0 \\
                              \end{array}\right),
\end{array}
\end{equation}
\begin{equation}\label{Eq:TBzzMatR2}
\begin{array}{l}
H^{\downarrow\uparrow}_{nn'}(k_x)=\lambda_\mathsmaller{\mathrm{R}}\left(\begin{array}{cccc}
                              0 & 0 & m'_4 & m'_5 \\
                              0 & 0 & m'_5 & m'_6 \\
                              m'_1 & m'_2 & 0 & 0 \\
                              m'_2 & m'_3 & 0 & 0 \\
                              \end{array}\right)
\end{array}
\end{equation}
with
\begin{equation}\label{Eq:TBzzMatRe}
\begin{array}{l}
m_{1/4}=-2\sqrt{3}\i\sin\left(\frac{k_xa}{2}\right)\delta_{n\pm1,n'},\\
m_{2/5}=\pm\left\{\delta_{n,n'}-\delta_{n\pm1,n'}\left[\cos\left(\frac{k_xa}{2}\right)\pm\sqrt{3}\sin\left(\frac{k_xa}{2}\right)\right]\right\},\\
m_{3/6}=\mp2\i\left[\delta_{n,n'}-\delta_{n\pm1,n'}\cos\left(\frac{k_xa}{2}\right)\right],\\
m'_{1/4}=2\sqrt{3}\i\sin\left(\frac{k_xa}{2}\right)\delta_{n\mp1,n'},\\
m'_{2/5}=\pm\left\{\delta_{n,n'}-\delta_{n\mp1,n'}\left[\cos\left(\frac{k_xa}{2}\right)\pm\sqrt{3}\sin\left(\frac{k_xa}{2}\right)\right]\right\},\\
m'_{3/6}=\pm2\i\left[\delta_{n,n'}-\delta_{n\mp1,n'}\cos\left(\frac{k_xa}{2}\right)\right]
\end{array}
\end{equation}
in the absence of magnetic fields.

At finite magnetic fields, Eqs.~(\ref{Eq:TBzzMatS})-(\ref{Eq:TBzzMatRe}) are modified by taking into account orbital effects via a Peierls phase. To preserve the good quantum number $k_x$, we use the Landau gauge $\bm{A}(\bm{r})=-B_z(y-W_y/2)\bm{e}_x$, where $W_y$ is the width of the nanoribbon in $y$-direction. Then, the Peierls phase due to $B_z$ results in a substitution of $k_xa$ by $k_xa-2\pi\phi(n-N_y/2)/\phi_0$ in $H^\mathrm{zz}_{nn'}(k_x)$. Here, $\phi=\sqrt{3}a^2B_z/2$ is the magnetic flux through one unit cell and $\phi_0=2\pi\hbar/e$ is the magnetic flux quantum. Moreover, a Zeeman term has to be added and we replace $H^\mathrm{zz}_{nn'}(k_x)$ in Eq.~(\ref{Eq:FullTBzz}) by $H^\mathrm{zz}_{nn'}(k_x)+H_\mathrm{Z}\delta_{n,n'}$, where $H_\mathrm{Z}$ is the matrix given in Eq.~(3) in the main text.

\begin{figure}[t]
\centering
\includegraphics*[width=8.5cm]{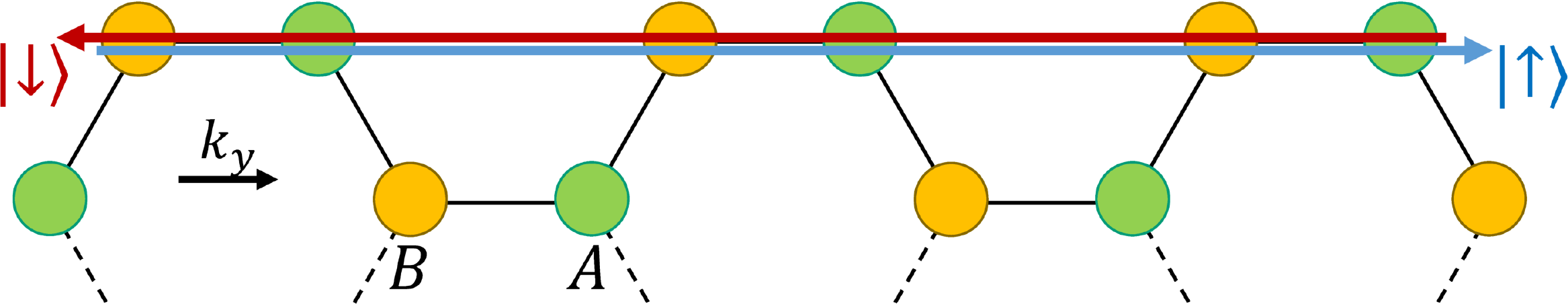}
\caption{Armchair nanoribbons with QSH edge states. Symbols $\uparrow/\downarrow$ refer to states which have predominantly spins $\uparrow/\downarrow$.}\label{fig:AC}
\end{figure}

\subsection{Armchair nanoribbons}
Similarly, we obtain nanoribbons with armchair (AC) edges along the $y$-direction (see Fig.~\ref{fig:AC}), a finite number of lattice sites $n=1,..,N_x$ along the $x$-direction, and a good momentum quantum number $k_y$ by a Fourier transformation with respect to $k_x/2$. The Hamiltonian of an AC nanoribbon then reads
\begin{equation}\label{Eq:FullTBac}
\hat{H}_\mathrm{ac}=\sum\limits_{k_y,n,n'}\hat{\Psi}^\dagger(n,k_y)H^\mathrm{ac}_{nn'}(k_y)\hat{\Psi}(n',k_y),
\end{equation}
where we have introduced field operators creating/annihilating electrons with momentum $k_y$ at lattice site $n$ in $x$-direction analogously to Eq.~(\ref{Eq:TBzzFO}). The corresponding matrix

\begin{equation}\label{Eq:TBacMat}
H^\mathrm{ac}_{nn'}(k_y)=\left(\begin{array}{cc}
                           H^{\uparrow\uparrow}_{nn'}(k_y) & H^{\uparrow\downarrow}_{nn'}(k_y) \\
                           H^{\downarrow\uparrow}_{nn'}(k_y) & H^{\downarrow\downarrow}_{nn'}(k_y) \\
                           \end{array}\right)
\end{equation}

contains
\begin{equation}\label{Eq:TBacMatS}
\begin{array}{l}
H^{\uparrow\uparrow/\downarrow\downarrow}_{nn'}(k_y)=\\
\quad\\
                           \left(\begin{array}{cccc}
                              0 & \mp\i\lambda_\mathsmaller{\mathrm{SOC}}\delta_{n,n'} & \tilde{h}_{xx} & \tilde{h}_{xy} \\
                              \pm\i\lambda_\mathsmaller{\mathrm{SOC}}\delta_{n,n'} & 0 & \tilde{h}_{xy} & \tilde{h}_{yy} \\
                              (\tilde{h}_{xx})^* & -(\tilde{h}_{xy})^* & 0 & \mp\i\lambda_\mathsmaller{\mathrm{SOC}}\delta_{n,n'} \\
                              -(\tilde{h}_{xy})^* & (\tilde{h}_{yy})^* & \pm\i\lambda_\mathsmaller{\mathrm{SOC}}\delta_{n,n'} & 0
                           \end{array}\right)
\end{array}
\end{equation}
with
\begin{equation}\label{Eq:TBacMatSe}
\begin{array}{l}
\tilde{h}_{xx}=V^1_\mathsmaller{pp\pi}\delta_{n,n'}+\frac{3V^1_\mathsmaller{pp\sigma}+V^1_\mathsmaller{pp\pi}}{4}\,\e^{\i\frac{\sqrt{3}k_ya}{2}}\left(\delta_{n+1,n'}+\delta_{n-1,n'}\right),\\
\tilde{h}_{xy}=\tilde{h}_{yx}=\frac{\sqrt{3}\left(V^1_\mathsmaller{pp\sigma}-V^1_\mathsmaller{pp\pi}\right)}{4}\,\e^{\i\frac{\sqrt{3}k_ya}{2}}\left(\delta_{n+1,n'}-\delta_{n-1,n'}\right),\\
\tilde{h}_{yy}=V^1_\mathsmaller{pp\sigma}\delta_{n,n'}+\frac{V^1_\mathsmaller{pp\sigma}+3V^1_\mathsmaller{pp\pi}}{4}\,\e^{\i\frac{\sqrt{3}k_ya}{2}}\left(\delta_{n+1,n'}+\delta_{n-1,n'}\right)
\end{array}
\end{equation}
and
\begin{equation}\label{Eq:TBacMatR1}
\begin{array}{l}
H^{\uparrow\downarrow}_{nn'}(k_y)=\lambda_\mathsmaller{\mathrm{R}}\left(\begin{array}{cccc}
                              0 & 0 & \tilde{m}_1 & \tilde{m}_2 \\
                              0 & 0 & \tilde{m}_2 & \tilde{m}_3 \\
                              \tilde{m}_4 & \tilde{m}_5 & 0 & 0 \\
                              \tilde{m}_5 & \tilde{m}_6 & 0 & 0 \\
                              \end{array}\right),
\end{array}
\end{equation}
\begin{equation}\label{Eq:TBacMatR2}
\begin{array}{l}
H^{\downarrow\uparrow}_{nn'}(k_y)=\lambda_\mathsmaller{\mathrm{R}}\left(\begin{array}{cccc}
                              0 & 0 & \tilde{m}'_4 & \tilde{m}'_5 \\
                              0 & 0 & \tilde{m}'_5 & \tilde{m}'_6 \\
                              \tilde{m}'_1 & \tilde{m}'_2 & 0 & 0 \\
                              \tilde{m}'_2 & \tilde{m}'_3 & 0 & 0 \\
                              \end{array}\right)
\end{array}
\end{equation}
with
\begin{equation}\label{Eq:TBacMatRe}
\begin{array}{l}
\tilde{m}_{1/4}=-\sqrt{3}\e^{\pm\i\frac{\sqrt{3}k_ya}{2}}\left(\delta_{n+1,n'}-\delta_{n-1,n'}\right),\\
\tilde{m}_{2/5}=\pm\left[\delta_{n,n'}-\e^{\pm\i\frac{\sqrt{3}k_ya}{2}}\frac{\left(1\mp\i\sqrt{3}\right)\delta_{n+1,n'}+\left(1\pm\i\sqrt{3}\right)\delta_{n-1,n'}}{2}\right],\\
\tilde{m}_{3/6}=\mp2\i\left(\delta_{n,n'}-\e^{\pm\i\frac{\sqrt{3}k_ya}{2}}\frac{\delta_{n+1,n'}+\delta_{n-1,n'}}{2}\right),\\
\tilde{m}'_{1/4}=\sqrt{3}\e^{\mp\i\frac{\sqrt{3}k_ya}{2}}\left(\delta_{n+1,n'}-\delta_{n-1,n'}\right)\\
\tilde{m}'_{2/5}=\pm\left[\delta_{n,n'}-\e^{\mp\i\frac{\sqrt{3}k_ya}{2}}\frac{\left(1\mp\i\sqrt{3}\right)\delta_{n+1,n'}+\left(1\pm\i\sqrt{3}\right)\delta_{n-1,n'}}{2}\right],\\
\tilde{m}'_{3/6}=\pm2\i\left(\delta_{n,n'}-\e^{\mp\i\frac{\sqrt{3}k_ya}{2}}\frac{\delta_{n+1,n'}+\delta_{n-1,n'}}{2}\right)
\end{array}
\end{equation}
in the absence of magnetic fields.

To account for finite magnetic fields and preserve the good quantum number $k_y$, we now use the Landau gauge $\bm{A}(\bm{r})=B_z(x-W_x/2)\bm{e}_y$, where $W_x$ is the width of the nanoribbon. Then, Eqs.~(\ref{Eq:TBacMatS})-(\ref{Eq:TBacMatRe}) are modified via a Peierls substitution $k_ya_\mathrm{ac}\to k_ya_\mathrm{ac}+2\pi\phi(n-N_x/2)/\phi_0$, where $\phi=\sqrt{3}a^2B_z/2$ and $\phi_0=2\pi\hbar/e$ as in the ZZ case above and $a_\mathrm{ac}=\sqrt{3}a$. Likewise, a Zeeman term has to be added and we replace $H^\mathrm{ac}_{nn'}(k_y)$ in Eq.~(\ref{Eq:FullTBac}) by $H^\mathrm{ac}_{nn'}(k_y)+H_\mathrm{Z}\delta_{n,n'}$, where $H_\mathrm{Z}$ is the matrix given in Eq.~(3) in the main text.
Diagonalizing the matrices $H^\mathrm{zz}_{nn'}(k_x)+H_\mathrm{Z}\delta_{n,n'}$ and $H^\mathrm{ac}_{nn'}(k_y)+H_\mathrm{Z}\delta_{n,n'}$ yields the spectra of ZZ and AC nanoribbons, respectively [see also Eq.~(\ref{Eq:TBSol}) below].

\begin{figure}[tb]                               
\begin{center}                                    
\includegraphics*[width=\linewidth]{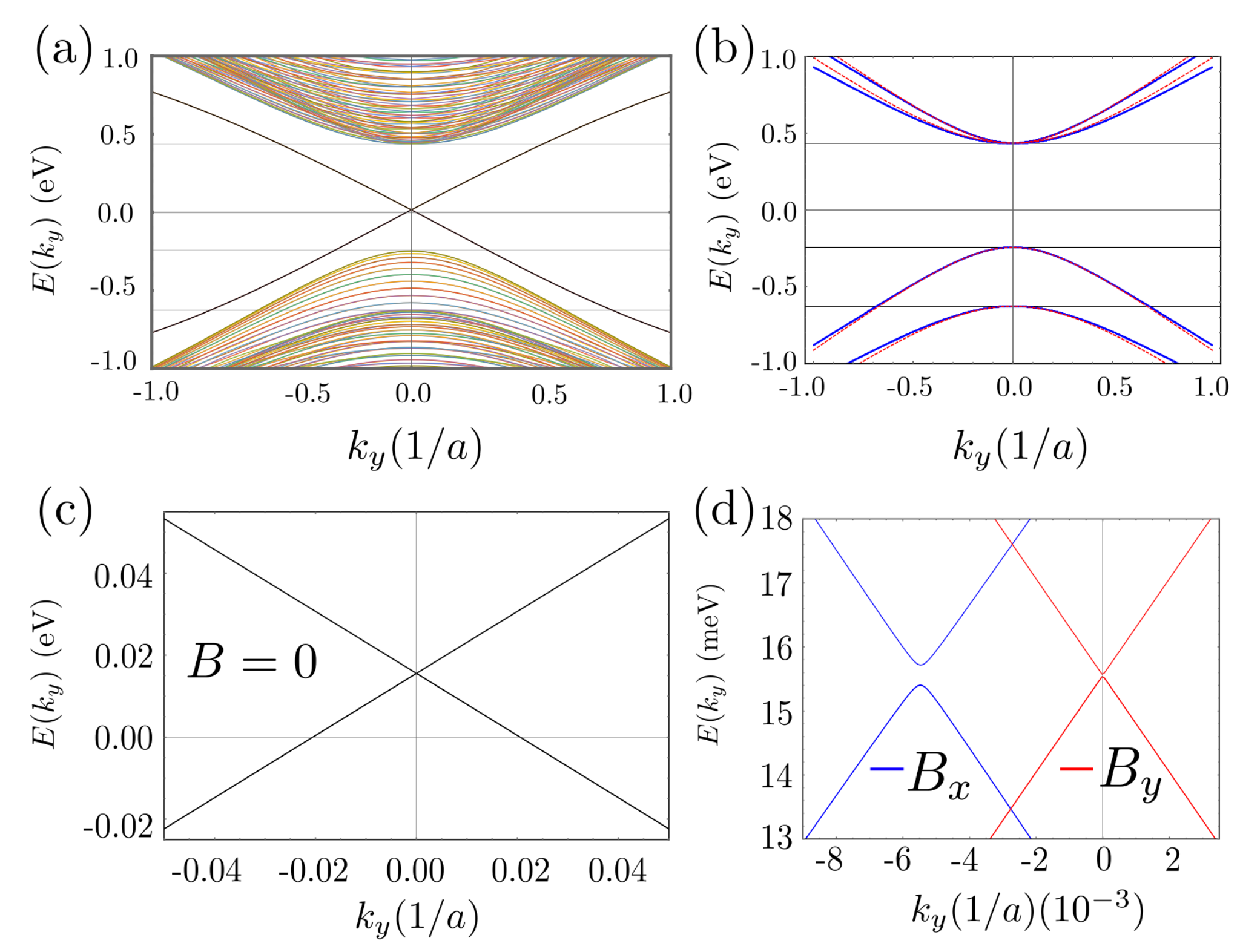}
\caption[]{
Panel (a) shows the energy dispersion of an AC nanoribbon with $N_y=100$ sites. In panel (b) we show the comparison of the bulk dispersion (solid) from Eq.~\eqref{Eq:TotalHamiltonian} and the linear expansion Eqs.~\eqref{effkm}-\eqref{rashbak1}, with $k_x=q_x=0$. 
Horizontal lines highlight the bulk gaps placed at $\lambda_\text{SOC}$, $-\lambda_\text{SOC}+6\lambda_\text{R}$ and $-\lambda_\text{SOC}-6\lambda_\text{R}$.
In panel (c), we zoom in the crossing of panel (a). Panel (d) shows a zoom into the crossing of panel (a) with the addition of a Zeeman term in $x$-direction (blue) and $y$-direction (red). In all panels we have used $\lambda_\text{SOC}=0.435\,$eV, $\lambda_\text{R}=0.032\,$eV, and in panel (d) $E_z^x=0.02\,$eV, $E_z^y=0.0\,$eV (blue) and $E_z^y=0.02\,$eV, $E_z^x=0.0\,$eV (red).
}  
\label{fig:Byrashba}                 
\end{center}
\end{figure}

\section{Low-energy Hamiltonian: Expansion around the \bf{k/K}' points}
We derive a low-energy Hamiltonian around the two inequivalent time-reversed $\bm{K}$ and $\bm{K}'$ points, placed at $\bm{K}=(-4\pi/3a,0)$ and $\bm{K}'=(4\pi/3a,0)$. 
To this aim, we first focus on the $4\times 4$ Hamiltonian $H_{\uparrow\uparrow,\downarrow\downarrow}$ given by Eq.~\eqref{Eq:HamiltonianDiagonalBlock} and rewrite it in the basis which diagonalizes the part of the Hamiltonian containing $\lambda_\mathrm{SOC}$, namely
\begin{widetext}
\begin{align}
\tilde{H}_{ss} (k_x,k_y)= 
\begin{pmatrix}
s\lambda_\mathrm{SOC}  & (\Delta_--i h_{xy}^{AB})^*   & 0 & \Delta_+^*  \\
\Delta_--i h_{xy}^{AB}    &  -s\lambda_\mathrm{SOC} & \Delta_+ & 0 \\
0& \Delta_+^* & -s\lambda_\mathrm{SOC}  & (\Delta_-+i h_{xy}^{AB})^*\\
\Delta_+ & 0  & \Delta_-+i h_{xy}^{AB}  & s\lambda_\mathrm{SOC}\\         
\end{pmatrix},
\label{eq:upnewbasis2}
\end{align} 
\end{widetext}
where $s=\uparrow/\downarrow$, and we have used $\Delta_\pm=1/2(h_{yy}\pm h_{xx})$ with the new basis 
$\{\Ket{i p_{x\sigma}^{B} +p_{y\sigma}^{B}},\Ket{-ip_{x\sigma}^{A}+p_{y\sigma}^{A}}$, $\Ket{-ip_{x\sigma}^{B}+p_{y\sigma}^B},\Ket{ip_{x\sigma}^{A}+p_{y\sigma}^A}\}$, denoted by the tilde.
The structure of the resulting Hamiltonian given in Eq.~\eqref{eq:upnewbasis2} is simple: It contains two block diagonal 2$\times$2 Hamiltonians coupled by an off-diagonal Hamiltonian proportional to $\Delta_+$. Close to the Dirac points, both blocks become effectively decoupled because to lowest order in $k$, $\Delta_+\approx 0$ and in addition the energy difference between the energies of both blocks is $\Delta E\approx3\,$eV. 
Thus, the low-energy physics in Eq.~\eqref{eq:upnewbasis2} is contained in the upper (lower) 2$\times$2 block for the $\bm{K}$ ($\bm{K}'$) point. Therefore, it is necessary to include the valley degree of freedom, yielding the linear Hamiltonian
\begin{align}
H_0=\hbar v_\text{F}(q_x \sigma_x \tau_z + q_y\sigma_y)+\lambda_{\text{SOC}} \sigma_z s_z \tau_z,
\label{effkm}
\end{align}
where $q_x$ and $q_y$ are measured from $\bm{K}/\bm{K}'$, and $\hbar v_\text{F}=\sqrt{3} a/4 (V_{pp\pi}^1-V_{pp\sigma}^1)$ is the Fermi velocity. Note that the resulting Hamiltonian has the same functional form as the low-energy expansion of the Kane-Mele model~\cite{Kane2005:PRL,*Kane2005:PRL2}. Here, we use $\sigma$, $s$, and $\tau$ as Pauli matrices representing the sublattice, spin, and valley degrees of freedom, respectively. In addition, there is an extra contribution coming from the Rashba SOC, given by $H_\text{R}= H_\text{R}^0+H_\text{R}^1$ where
\begin{align}
&H_\text{R}^0=3\lambda_\text{R} (\sigma_x s_y \tau_z-\sigma_y s_x),\label{rashbak0}\\
&H_\text{R}^1=\sqrt{3}\lambda_\text{R}q_y a(\sigma_x s_x +\sigma_y s_y\tau_z),\label{rashbak1}
\end{align}
where $a$ is the lattice constant, introduced previously.
It is interesting to note that close to the $\bm{K/K'}$ points, the Rashba contribution mixes different spins and different $A$-$B$ sublattices only within the valence band. 
At $q_y=0$, the term Eq.~\eqref{rashbak0} opens a gap within the valence band of $-\lambda_\text{SOC}\pm 6\lambda_\text{R}$, as can be observed in Figs.~\ref{fig:Byrashba}(a,b). 
In addition, in Fig.~\ref{fig:Byrashba}(b) we provide a comparison between the bulk eigenenergies of the linear expansion [Eqs.~\eqref{effkm}-\eqref{rashbak1}], and the 
full Hamiltonian [Eq.~\eqref{Eq:TotalHamiltonian}]. In both cases we use $k_x=q_x=0$, and obtain an almost perfect matching for $q_y a< 0.5$.

\section{Symmetry analysis}

\subsection{Bulk symmetries}

In the absence of Rashba SOC, the bulk Hamiltonian given by Eq.~\eqref{Eq:TotalHamiltonian} 
exhibits time-reversal ($\mathcal{T}$), particle-hole ($\mathcal{C}$) and chiral symmetry ($\mathcal{S}$) 
in the whole Brillouin zone. 
The specific forms of these operators are 
\begin{align}
&\mathcal{T}= -\i \pi_0 \overline{\sigma}_0 s_y \mathcal{K},\label{time}\\
&\mathcal{C}= \pi_0 \overline{\sigma}_z s_{0/z}\label{partichole}\mathcal{K},\\
&\mathcal{S}=\mathcal{T} \mathcal{C}=-\i   \pi_0 \overline{\sigma}_z s_{x/y}\label{chiral}.
\end{align}
Here, $\pi$, $\overline{\sigma}$ and $s$ describe the orbital $p_x/p_y$, sublattice and spin subspaces, respectively.
Furthermore, $\mathcal{K}$ is the complex conjugation operator.

For further purposes, it is convenient to express these symmetries around the $\bm{K}/\bm{K}'$ points, where the AC edge states cross.
To this aim, we perform the same change of basis as in Eq.~\eqref{eq:upnewbasis2}, namely
\begin{align}
&\mathcal{T}_\text{K/K'}= -\i \sigma_0 s_y \tau_x\mathcal{K},\label{time2}\\
&\mathcal{C}_\text{K/K'}= \sigma_z s_{0/z} \tau_x \mathcal{K},\label{partichole2}\\
&\mathcal{S}_\text{K/K'}=-\i   \sigma_z s_{x/y} \tau_0 \label{chiral2}.
\end{align}
Here, the $\sigma$ matrix refers also to the $A$ and $B$ sublattices. However, it involves a specific combination of the $p_x/p_y$ orbitals (see above). 
In order to highlight this difference, we have removed the bar from $\overline{\sigma}$, that is,~$\overline{\sigma}\to \sigma$.  
We recall that the Pauli matrices $\tau$ refer to the $\bm{K}/\bm{K}'$ points, similar to the low-energy description of graphene.

\subsection{Lattice symmetries}

In the absence of Rashba SOC ($\lambda_\text{R}= 0$) the bulk Hamiltonian given by Eq.~\eqref{Eq:TotalHamiltonian} exhibits inversion and reflection symmetries. 
In the full basis, inversion symmetry reads $\mathcal{I}=\pi_0 \overline{\sigma}_x s_0$ and 
transforms the Hamiltonian as
\begin{align}
\mathcal{I} H(k_x,k_y) \mathcal{I}^{-1}=H(-k_x,-k_y).
\end{align}
In addition, $\mathcal{I}$ can be decomposed into the reflection symmetries 
$\mathcal{R}(x)$ and $\mathcal{R}(y)$, which transform the Hamiltonian
as 
\begin{align}
& \mathcal{R}(x) H(k_x,k_y) \mathcal{R}(x)^{-1}=H(-k_x,k_y),\\
& \mathcal{R}(y) H(k_x,k_y) \mathcal{R}(y)^{-1}=H(k_x,-k_y),
\end{align}
and their explicit form is given by
\begin{align}
& \mathcal{R}(x)=\pi_z \overline{\sigma}_0 s_{x/y},\\
& \mathcal{R}(y) = \pi_z \overline{\sigma}_x s_{x/y},
\end{align}
fulfilling  $\mathcal{I}=\mathcal{R}(x)\mathcal{R}(y)$.

In addition, the low-energy Hamiltonian fulfills the following symmetries
\begin{align}
& \mathcal{R}_\text{K/K'}(x) = \sigma_0 s_{x/y} \tau_x,\\
& \mathcal{R}_\text{K/K'}(y)=\sigma_x s_{x/y} \tau_0, \\
& \mathcal{I}_\text{K/K'} =\sigma_x s_{0} \tau_x.
\end{align}

\subsection{Armchair and zigzag boundary conditions}

The boundary conditions for AC and ZZ nanoribbons impose 
\begin{align}
&\Psi(0,y)=M_\text{ac} \Psi(0,y),\label{bcac}\\
&\Psi(x,0)=M_\text{zz} \Psi(x,0),\label{bczz} 
\end{align}
respectively.
Here, $M_\text{zz/ac}$ are obtained from canceling the current perpendicular to the edge $\langle I_\perp^\text{zz/ac}\rangle $ \cite{Brey2006a, Akhmerov2008a}. This condition 
can be expressed as $\{M_\text{zz/ac},I_\perp^\text{zz/ac}\}=0$.
Knowing that close to the high-symmetry $\Gamma$ and $\bm{K}/\bm{K}'$ points, 
the perpendicular current is given by
$I_\perp^\text{zz}\propto \overline{\sigma}_y s_0$, and $I_\perp^\text{ac}= \sigma_x s_0 \tau_z$, respectively, we obtain 
\begin{align}
&M_\text{ac}=\sigma_0 s_0 \tau_x,\label{mac}\\ 
&M_\text{zz}=\pi_0 \overline{\sigma}_z s_0.\label{mzz}
\end{align}
Note that $M_\text{ac}$ and $M_\text{zz}$ commute with the bulk symmetries given by 
Eqs.~\eqref{time}-\eqref{chiral} and Eqs.~\eqref{time2}-\eqref{chiral2}.

%

\subsection{Helicity operators at the crossing points for $\lambda_\text{R}=0$}

We now use the derived symmetries and find the helicity operators $\mathcal{O}_\text{zz/ac}$ at the crossing points. 
To this aim, we will use the symmetry relations on the edge wave functions $\Psi_{\pm,\mathsmaller{E}}(q)$. 
Here, $q$ is the momentum along the edge with respect to the crossing point, and $\pm$ denotes the branch solution.
In addition, we use a redundant label $E$ to denote the energy of the wave function. 
This will be helpful when applying different symmetries. However, we will drop this label everywhere else.

We start from chiral symmetry, which relates two branches $\pm$ at a given $q$, namely
\begin{align}
\Psi_{\pm,\mathsmaller{E}}(q)=\pm \mathcal{S} \Psi_{\mp,-\mathsmaller{E}}(q).
\end{align}
In addition to $\mathcal{S}$, the reflection symmetry $\mathcal{R}$ 
relates the counter-propagating modes $q\to -q$,
\begin{align}
\Psi_{\pm,\mathsmaller{E}}(q)=\mathcal{R} \Psi_{\mp,\mathsmaller{E}}(-q),
\end{align}
Thus, at the crossing point $q=0$ and $E=0$, 
the action of $\mathcal{R}$ and $\mathcal{S}$
leads to
\begin{align}
\Psi_{\pm,E=0}(q=0)=\pm\mathcal{O}\Psi_{\pm,E=0}(q=0),
\end{align}
where the helicity operator $\mathcal{O}=\mathcal{R} \mathcal{S}$ determines the symmetry of the wave function. 
Note that here we need to select a $\mathcal{R}$ operator that anticommutes with $\mathcal{S}$. 
Otherwise, the helicity operator would commute with $\mathcal{S}$ and does not provide orthogonal Kramers partners. For this reason, we select $\mathcal{R}(y)=\sigma_x s_y\tau_0$ and $\mathcal{R}(x)=\pi_z \overline{\sigma}_0 s_x$, yielding the helicity operators 
\begin{align}
&\mathcal{O}_\text{zz}=\mathcal{R}(x) \mathcal{S}=-\i\pi_z \overline{\sigma}_z s_z,\\
&\mathcal{O}_\text{ac}= \mathcal{R}(y) \mathcal{S}_\text{K/K'}= -\i\sigma_y s_0 \tau_0.
\end{align}

\begin{figure}[tb]                               
\begin{center}                                    
\includegraphics*[width=8.5cm]{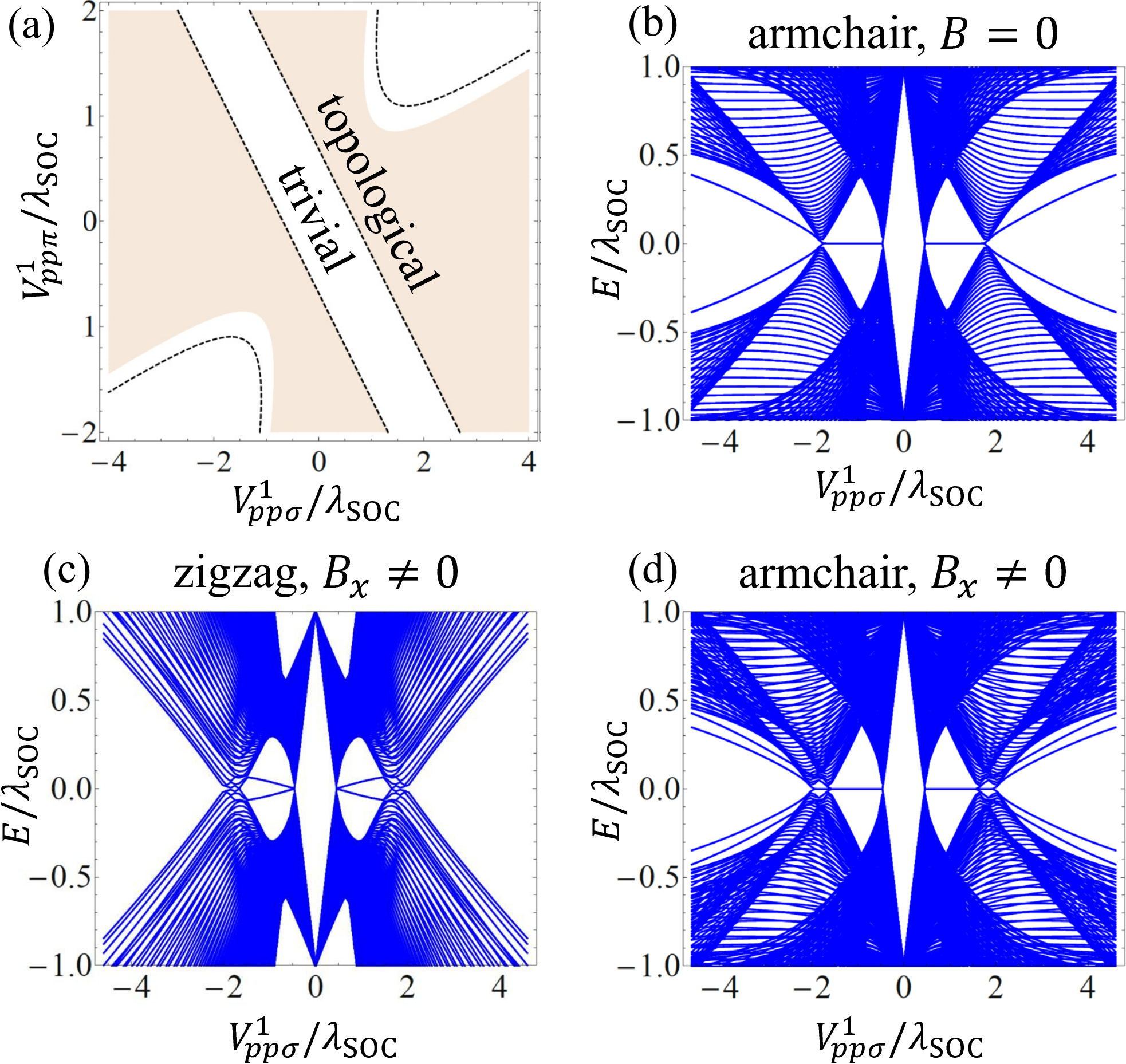}
\caption{(a) Phase diagram for $B=0$ according to the topological invariant given in Eqs.~\eqref{topoinv}-\eqref{delta2}. The topological invariant is plotted as a function of $V_{pp\sigma}^1$ and $V_{pp\pi}^1$ in units of $\lambda_\text{SOC}$. Lighter (darker) areas denote topological (trivial) regimes. The black lines in panel~(a) indicate the shifted boundaries of the topological regime given by Eqs.~\eqref{eq:TIZeeman}-\eqref{eq:PfZeeman2} in the presence of an in-plane Zeeman term $B_\parallel=0.23\lambda_\text{SOC}$. (b)-(d) Dispersion of nanoribbons for fixed momentum $k_{x/y}=0$, $\lambda_\text{SOC}=435$ meV, and $\lambda_\text{R}=0$ as a function of $V_{pp\sigma}^1$ along the line $V_{pp\pi}^1=0.5V_{pp\sigma}^1$ [that is, the diagonal in panel~(a)]: (b) AC nanoribbons for $B=0$, (c) ZZ nanoribbons for $B_\parallel=100$ meV$=0.23\lambda_\text{SOC}$, (d) AC nanoribbons for $B_\parallel=100$ meV$=0.23\lambda_\text{SOC}$.}  
\label{fig:phasediagram}                 
\end{center}
\end{figure}

\section{Topological invariants in the presence of spatial symmetries}

In this section, we calculate the topological invariant of the bismuthene Hamiltonian in the absence of Rashba SOC ($\lambda_\text{R}=0$). We will do this, both in the presence and absence of an in-plane Zeeman term responsible for breaking TRS and chiral symmetry. 

\subsection{Absence of magnetic field:  DIII symmetry class}

The combination of time-reversal, particle-hole and chiral symmetries sets the Hamiltonian in symmetry class DIII, with the topological invariant $n_{z_2}$. In order to calculate it, we take advantage of the presence of inversion symmetry $\mathcal{I}$. In this situation, we can calculate the topological invariant as \cite{Fu2007a}
\begin{align}
&(-1)^{n_{z_2}}=\prod_{i=TRIM} \delta_i,\label{topoinv}\\
&\delta_i=\prod_{m=1}^n \xi_{2m} (\Gamma_i)\label{deltas},
\end{align}
where $\xi_{2m} (\Gamma_i)=\pm 1$ is the parity eigenvalue of the 2$m$th occupied energy band evaluated at the $i$th-time reversal invariant momentum (TRIM) $\Gamma_i$. 
Note that the product only takes into account the half of the Kramers partners from the $n$ occupied states. The TRIM are  
\begin{align}
&\Gamma_1=(0,0),\\
&\Gamma_2=\frac{2\pi}{a}(0,\frac{1}{\sqrt{3}}),\\
&\Gamma_3=\frac{2\pi}{a}(\frac{1}{2},-\frac{1}{2\sqrt{3}}),\\
&\Gamma_4=\frac{2\pi}{a}(-\frac{1}{2},-\frac{1}{2\sqrt{3}}),
\end{align}
which correspond to the $\Gamma$-point and one of the $M$-points highlighted in Fig.~1 in the main text. An easy way to find the topological invariant consists of rewriting 
the Hamiltonian in the basis which diagonalizes the inversion symmetry $\mathcal{I}$. 
Since $[H(\Gamma_i),\mathcal{I}]=0$, the resulting matrix is block diagonal, 
where each block exhibits a different parity eigenvalue  $\xi_{2m} (\Gamma_i)=\pm 1$.

Substituting the TRIM into Eq.~\eqref{deltas}, we find that $\delta_3=\delta_4$, and therefore,
\begin{align}
&\delta_1=\text{Sign}\left[ \frac{9}{4}(V_{pp\pi}^1+V_{pp\sigma}^1)^2-\lambda_\text{SOC}^2   \right],\\ 
&\delta_2=\text{Sign}\left[  \frac{1}{4}(V_{pp\pi}^1+V_{pp\sigma}^1)^2-\lambda_\text{SOC}^2 -(V_{pp\pi}^1-V_{pp\sigma}^1)^2 \right]\label{delta2}
\end{align}
determine whether the Hamiltonian is trivial or topological. In Fig.~\ref{fig:phasediagram}(a), we show the numerical evaluation of the topological invariant $n_{z_2}$. Taking into account that $|\lambda_\text{SOC}|>|V_{pp\pi}^1|$, the topological region becomes present for $|V_{pp\sigma}^1|>2/3 \lambda_\text{SOC}$, which sets the central diagonal stripe in Fig.~\ref{fig:phasediagram}(a).

To illustrate the validity of the phase diagram presented in Fig.~\ref{fig:phasediagram}(a), Fig.~\ref{fig:phasediagram}(b) shows the particle-hole symmetric energy spectrum of an AC nanoribbon with $\lambda_\text{SOC}=435$ meV, $\lambda_\text{R}=0$ and zero magnetic field ($B=0$) at fixed momentum $k_y=0$. The spectrum is plotted as a function of $V_{pp\sigma}^1$ along the line $V_{pp\pi}^1=0.5V_{pp\sigma}^1$, that is, along the diagonal in Fig.~\ref{fig:phasediagram}(a). If $V_{pp\pi}^1$ and $V_{pp\sigma}^1$ are situated in the topological regime, the spectrum in Fig.~\ref{fig:phasediagram}(b) exhibits a flat line of zero-energy states. This flat line corresponds to the crossing of the two counter-propagating, gapless edge states at $k_y=0$ in the QSH regime. In the trivial regime, a gap is opened at $k_y=0$.

\subsection{Presence of magnetic field:  D Symmetry class}

The presence of an in-plane Zeeman term $H_\text{B} = B_\parallel(\cos(\theta)s_x+\sin(\theta)s_y)$ breaks time-reversal and chiral symmetries, 
placing the Hamiltonian given by Eqs.~\eqref{Eq:TotalHamiltonian}-\eqref{Eq:HamiltonianDiagonalBlockSlaterKoster} in symmetry class D with the topological invariant $Z$, the winding number. However, the presence of reflection symmetries can modify the topological invariant, giving rise to what is known as crystalline topological insulators \cite{Fu2011a, Chiu2013a, Chiu2016a}.
In order to classify the Hamiltonian accounting for reflection symmetries, we have to compute the commutation relations of $\mathcal{R}(x)$ and $\mathcal{R}(y)$ with the operator describing PHS, $\mathcal{C}=-\i \pi_0 \overline{\sigma}_z s_{0/z}$~\cite{Chiu2013a, Chiu2016a}. We find that
\begin{align}
&\{\mathcal{R}(x),\mathcal{C}\}=0, \\
& [\mathcal{R}(y),\mathcal{C}]=0. 
\end{align}
%
Since $\mathcal{R}(x)$ anticommutes with $\mathcal{C}$, it can be proven that in 2D there is always a trivial phase.
In turn, $\mathcal{R}(y)$, commutes with $\mathcal{C}$ and in 2D giving rise to the topological number $MZ_2$, the mirror Chern number~\cite{Chiu2013a, Chiu2016a}. 

The idea now is to project the 2D Hamiltonian on the 1D reflection invariant momenta (RIM) and calculate the topological invariant of the resulting 1D Hamiltonians.
The RIM are given by $k_y=0$ and $k_y=2\pi/(\sqrt{3}a)$, which have the property $H(k_x,k_y)=H(k_x,-k_y)$.
At these points, the effective 1D Hamiltonians commute with $\mathcal{R}(y)$, that is, $[H_{k_y}(k_x),\mathcal{R}(y)]=0$. 
Therefore, it is possible to use the same basis that diagonalizes $R(y)$, that is, $U_R \mathcal{R}(y)U_R^\dagger=\text{diag}({\bf 1}_{4\times 4},-{\bf 1}_{4\times 4})$, to rewrite $H(k_x,k_y)$ in a block diagonal basis, that is, $U_R H_{k_y}(k_x) U_R^\dagger=\text{diag}(H^+_{k_y}(k_x),H^-_{k_y}(k_x))$, with the two reflection parity ($\pm$) blocks given by
\begin{widetext}
\begin{align}
H^{\pm}_{k_y}(k_x) = 
\begin{pmatrix}
0  & -\i \lambda_\text{SOC}   & \mp B_\parallel+h_{yy}^\text{AB}& h_{xy}^\text{AB} \\
\i \lambda_\text{SOC}  & 0   & h_{xy}^\text{AB} & \pm B_\parallel+h_{xx}^\text{AB}  \\
\mp B_\parallel+h_{yy}^\text{AB}  & h_{xy}^\text{AB}   & 0 & -\i \lambda_\text{SOC}   \\
h_{xy}^\text{AB}&  \pm B_\parallel+h_{xx}^\text{AB}    & \i \lambda_\text{SOC} & 0    \\         
\end{pmatrix},
\label{eq:Hrefl}
\end{align} 
where we have used the fact that at the RIM points, the Hamiltonian elements $h_{xx}^\text{AB},h_{yy}^\text{AB},h_{xy}^\text{AB}\in \mathbb{R} $.

The topological invariant of the resulting 1D {\bf D}-class Hamiltonian can be calculated as in the Kitaev model, see Ref.~\onlinecite{Kitaev2001:PhysUs}.
To calculate it, we express Eq.~\eqref{eq:Hrefl} in a basis (Majorana basis) in which the unitary part of the particle-hole operator $\mathcal{C}=U_c \mathcal{K}$ transforms into 
$U_c={\bf 1}$. 
At the particle-hole invariant momenta (PHIM), the Hamiltonian becomes purely imaginary $H_M=i A$, where $A$ is a real and antisymmetric matrix, $A^T=-A$ given by
\begin{align}
A^\pm= 
\begin{pmatrix}
0  & - \lambda_\text{SOC}   & \mp B_\parallel+h_{yy}^\text{AB}& 0 \\
\lambda_\text{SOC}  & 0   & 0 & \pm B_\parallel+h_{xx}^\text{AB}  \\
-(\mp B_\parallel+h_{yy}^\text{AB})  & 0   & 0 & - \lambda_\text{SOC}   \\
0 &  -(\pm B_\parallel+h_{xx}^\text{AB})    &  \lambda_\text{SOC} & 0    \\         
\end{pmatrix}.
\label{eq:Hreflmajo}
\end{align} 
This can be understood from the particle-hole 
transformation $U_c H(-k_x,-k_y)^* U_c^\dagger=-H(k_x,k_y)$. 
Then, at these points the invariant is expressed in terms of the Pfaffian, 
\begin{align}\label{eq:TIZeeman}
(-1)^{n_{z_2}^\pm} =\text{Sign}\{ \text{Pf}[ A^\pm_{k_y=0}(0)]\} \text{Sign}\{ \text{Pf}[ A^\pm_{k_y=\frac{2\pi}{\sqrt{3}a}}(0)]\}, 
\end{align}
with 
\begin{align}
&\text{Pf}[ A^\pm_{k_y=0}(0)]= \lambda^2_\text{SOC} + B_\parallel^2-\frac{9}{4}(V_{pp\pi}^1+V_{pp\sigma}^1)^2,\label{eq:PfZeeman}\\
&\text{Pf}[ A^\pm_{k_y=\frac{2\pi}{\sqrt{3}a}}(0)]= \lambda^2_\text{SOC} + \frac{1}{4}\left(\mp 2B_\parallel + V_{pp\pi}^1-3V_{pp\sigma}^1\right) \left(\pm 2B_\parallel + 3V_{pp\pi}^1-V_{pp\sigma}^1\right).\label{eq:PfZeeman2}
\end{align}
\end{widetext}
The topological invariant is calculated as 
\begin{align}\label{eq:TIZeeman2}
(-1)^{n_{z_2}^\pm} =\text{Sign}\{ \text{Pf}[ A_{k_y=0}(0)]\} \text{Sign}\{ \text{Pf}[A^\pm_{k_y=\frac{2\pi}{\sqrt{3}a}}(0)]\}.
\end{align}
As long as the bulk gap remains finite, the topological invariants $n_{z_2}^\pm$ exhibit the same value. When the bulk gap closes, the topological classification becomes modified and $MZ_2$ does no longer hold turning into $MZ$, see Ref.~\cite{Chiu2014:PRB}. Further details about the topological classification will be presented elsewhere.
%
%
Due to the small energy scale provided by the Zeeman energy compared to $\lambda_\text{SOC}$, $V_{pp\sigma}^1$ and $V_{pp\pi}^1$, 
the presence of a finite $B_\parallel$ does not change significantly the position of the topological transition obtained in the $B_\parallel=0$ case, see Eqs.~\eqref{topoinv}-\eqref{delta2}.
An example of the phase diagram with finite magnetic field given by Eqs.~\eqref{eq:TIZeeman}-\eqref{eq:PfZeeman2} is shown in Fig.~\ref{fig:phasediagram}(a) 
for an in-plane Zeeman term $B_\parallel=0.23\lambda_\text{SOC}$. There, the black lines indicate the boundaries of the topological 
regime and how these boundaries are shifted compared to the QSH phase at $B=0$ (colored surfaces). 

It is important to remark at this point that until now we were discussing the symmetries and topological invariant of the bulk Hamiltonian, and therefore, these arguments apply in principle 
to both ZZ and AC boundary conditions. 
In order to understand the differences observed in the main text, one has to realize that ZZ boundary conditions do not commute with the reflection symmetry 
$\mathcal{R}(y)$ [see Eq.~\eqref{mzz}]. Thus, ZZ nanoribbons in the presence of magnetic fields always exhibit a trivial phase adopted from $\mathcal{R}(x)$. 
In turn, AC boundary conditions preserve the reflection symmetry $\mathcal{R}(y)$, yielding a topological number $n_{z_2}$. 
This is corroborated by Figs.~\ref{fig:phasediagram}(c) and~(d), which show the energy spectra of (c) ZZ and (d) AC nanoribbons for the same parameters as in Fig.~\ref{fig:phasediagram}(b), but with finite $B_\parallel=0.23\lambda_\text{SOC}=100$ meV: Plotted as a function of $V_{pp\sigma}^1$ along the line $V_{pp\pi}^1=0.5V_{pp\sigma}^1$, only AC nanoribbons exhibit gapless edge states if $V_{pp\sigma}^1$ and $V_{pp\pi}^1$ 
are situated in the topological regime [see Fig.~\ref{fig:phasediagram}(d)]. In ZZ nanoribbons, on the other hand, a gap is always opened by $B_\parallel$, as shown in Fig.~\ref{fig:phasediagram}(c).

The robustness of AC QSH edge states even if TRS is broken by a finite in-plane magnetic field is in certain ways reminiscent of the situation in chiral semimetals: Here, edge states survive even if chiral symmetry is partially broken by the boundary conditions or extra terms in the Hamiltonian~\cite{Kharitonov2017:PRL}.

\section{Armchair edge states}
Once we have established the low-energy Hamiltonian given in Eqs.~\eqref{effkm}--\eqref{rashbak1}, we can derive an analytical expression for the edge states with the AC boundary condition $\Psi(0,y)=M_\text{ac}\Psi(0,y)$, with $M_\text{ac}$ given in Eq.~\eqref{mac}. 
Besides, we analyze a semi-infinite plane defined for $x>0$, and therefore we look for exponentially decaying solutions such that $\psi_k(\infty)=\psi_{k'}(\infty)=0$. 
An analytical solution to Eq.~\eqref{effkm} with AC boundary conditions was first derived in Ref.~\onlinecite{Prada2011a}. However, in Ref.~\onlinecite{Prada2011a} the symmetries discussed in the previous section were not analyzed and finite $\lambda_\text{R}$ or $\bm{B}$ were also not taken into account. As we expected, the results of Ref.~\onlinecite{Prada2011a} coincide with the helicity operator derived in the previous section. Here, we derive a more general case with $\lambda_\text{R}\neq 0$. We first consider $H_\text{R}^0$, that is, the $q_y$-independent contribution given by Eq.~\eqref{rashbak0}, and then we add perturbatively $H_\text{R}^1$, that is, the $q_y$-dependent one given by Eq.~\eqref{rashbak1}. 

\subsection{General solution}

Taking into account the AC boundary condition implies a mixing of the two valleys, leading to the general ansatz
\begin{widetext}
\begin{align}
\Psi(x,y)=\left[
\left(\alpha_1 e^{-\kappa_1 x}
\varphi_{1,k}+
 \alpha_2 e^{-\kappa_2 x}
\varphi_{2,k} \right)
 e^{i K_x x}+
 \left(
 \beta_1 e^{-\kappa_1 x}
 \varphi_{1,k'}
 +
 \beta_2 e^{-\kappa_2 x}
 \varphi_{2,k'}
 \right)
 e^{i K_x' x}\right] e^{i q_y y},
 \label{ansatz}
\end{align}
where $K_x=-K_x'=4\pi/3a$.
Then, we apply the Hamiltonian from Eqs.~\eqref{effkm}--\eqref{rashbak0} with 
$k_{x,y}\rightarrow -i\partial_{x,y}$ to Eq.~\eqref{ansatz}, finding 
\begin{align}
&\kappa_1= \frac{1}{\hbar v_\text{F}}\sqrt{(\hbar v_\text{F} q_y)^2+\lambda_{\text{SOC}}^2-6\lambda_{\text{R}}\lambda_{\text{SOC}}+6\lambda_{\text{SOC}}E-E^2}, \\
&\kappa_2= \frac{1}{\hbar v_\text{F}}\sqrt{(\hbar v_\text{F} q_y)^2+\lambda_{\text{SOC}}^2+6\lambda_{\text{R}}\lambda_{\text{SOC}}-6\lambda_{\text{SOC}}E-E^2}.
\end{align}
Next, we diagonalize the Hamiltonian with $\kappa=\kappa_1$ and $\kappa=\kappa_2$ for both valleys, yielding
\begin{align}
\begin{pmatrix}
\varphi_{1,k}, & \varphi_{2,k}, & \varphi_{1,k'}, & \varphi_{2,k'}\\
\end{pmatrix}=
\begin{pmatrix}
\begin{pmatrix}
a_1\\
b_1\\
i b_1\\
c_1 \\
\end{pmatrix},
\begin{pmatrix}
a_2\\
-b_2\\
-i b_2\\
c_2\\
\end{pmatrix},
\begin{pmatrix}
i b_1\\
-i a_1\\
-i c_1\\
b_1\\
\end{pmatrix},
\begin{pmatrix}
i b_2\\
-i a_2\\
i c_2\\
-b_2\\
\end{pmatrix}
\end{pmatrix},
\label{eig_rashba}
\end{align}
\end{widetext}
where
\begin{align}
&a_j= i(-q_y+\kappa_j),\\
&b_j= (E-\lambda_\text{SOC})/\hbar v_\text{F},\\
&c_j= q_y+\kappa_j.
\end{align}
We now impose the boundary condition from Eq.~\eqref{bcac} by canceling the determinant composed of the vectors of Eq.~\eqref{eig_rashba} with $x=0$. 
The resulting equation allows us to obtain the solutions for $E$. In general, the eigenenergies have cumbersome expressions. However, close to $q_y\approx 0$ the expressions simplify considerably:
\begin{align}
E_\pm\approx\frac{9 \lambda_\text{R}^2}{\lambda_\text{SOC} }\pm \hbar v_\text{F} q_y \left(1-\frac{9 \lambda_\text{R}^2}{\lambda_\text{SOC} ^2}\right)
\end{align}
and
\begin{align}
&\kappa_1\approx \frac{1}{\hbar v_\text{F}}(\lambda_\text{SOC}-3\lambda_\text{R} + \hbar v_\text{F} q_y \frac{3\lambda_\text{R}}{\lambda_\text{SOC}}),\\
&\kappa_2\approx \frac{1}{\hbar v_\text{F}}(\lambda_\text{SOC}+3\lambda_\text{R} - \hbar v_\text{F} q_y \frac{3\lambda_\text{R}}{\lambda_\text{SOC}}).
\end{align}

\subsection{Edge states close to $q_y=0$}

At $q_y=0$, it is possible to obtain an analytical expression for eigenstates fulfilling the boundary condition given in Eq.~\eqref{bcac}. 
At this point, the two eigenstates cross at $E_{\pm}=9\lambda_\text{R}^2/\lambda_\text{SOC}$ and the eigenvectors simplify considerably since $\varphi_{1,k} =\varphi_{2,k'}$ and $\varphi_{2,k} =\varphi_{1,k'}$, yielding
\begin{align}
&\Psi_{1}(x)=\frac{(e^{-\kappa_1 x}e^{i K_x x}-e^{i K_x' x}e^{-\kappa_2 x})}{2\sqrt{\lambda_\text{SOC}}N}
\begin{pmatrix}
-i \lambda_-\\
 \lambda_+\\
 i\lambda_+ \\
 \lambda_- \\
\end{pmatrix},
  \label{solrashbaup}\\
&\Psi_{2}(x)=\frac{(e^{-\kappa_2 x}e^{i K_x x}-e^{i K_x' x}e^{-\kappa_1 x})}{2\sqrt{\lambda_\text{SOC}}N}
\begin{pmatrix}
i\lambda_+\\
 -\lambda_- \\
 i\lambda_- \\
 \lambda_+\\
\end{pmatrix},
  \label{solrashbadown}
\end{align}
where we used $\lambda_\pm=\sqrt{\lambda_\text{SOC}\pm 3\lambda_\text{R}}$ and $N$ is the normalization factor
\begin{align}
N^2=\frac{1}{2}\left(\frac{1}{\kappa_1}+\frac{1}{\kappa_2}-4\frac{\kappa_1+\kappa_2}{(\kappa_1+\kappa_2)^2+4K_x^2}\right).
\end{align}
After adding the perturbation $V=\hbar v_\text{F} q_y \sigma_y$, we obtain the energy dispersion 
\begin{align}
E_\pm=\frac{9\lambda_\text{R}^2}{\lambda_\text{SOC}}\pm \hbar v_\text{F} q_y,
\end{align}
with the eigenstates 
\begin{align}
\Psi_{\pm}(x,q_y)=\frac{1}{\sqrt{2}} \left( \Psi_1(x)\pm\Psi_2 (x) \right)e^{\i q_y y}.
\label{estatesr}
\end{align}
As we explained above, both $H_\text{R}^0$ and $H_\text{R}^1$ break chiral symmetry $\mathcal{S}_\text{K/K'}$, 
and in principle the helicity operator present for $\lambda_\text{R}=0$, that is, $\mathcal{O}_\text{ac}=-\i \sigma_y s_0 \tau_0$, should no longer hold.
However, we can check that the eigenstates $\Psi_\pm$, resulting from $H_\text{R}^0$ still fulfill $\langle \Psi_\pm|\sigma_y s_0|\Psi_\pm \rangle=\pm 1$. 
This means that $H_\text{R}^0$ breaks only slightly chiral symmetry, yielding a crossing point protected against an applied magnetic field.
Indeed, we can observe this by adding perturbatively an in-plane magnetic field
$H_\text{B}= E_Z^x s_x+ E_Z^y s_y$, leading to the energy dispersion
\begin{align}
E_\pm \approx \frac{9 \lambda_\text{R}^2}{\lambda_\text{SOC} }\pm  \left( \hbar v_\text{F} q_y + E_Z^x \frac{3\lambda_\text{R}}{\lambda_\text{SOC}} \right).
\label{pertdisp}
\end{align}
Here, the crossing point does not open, but shifts towards
\begin{align}
q_y=  -\frac{E_Z^x}{\hbar v_\text{F}} \frac{3\lambda_\text{R}}{\lambda_\text{SOC}}.
\label{shift}
\end{align}
This shift is confirmed numerically in Fig.~\ref{fig:Byrashba}(d). 
Note however, that the term proportional to $E_Z^y$ does not participate in the shift in Eq.~\eqref{pertdisp}. 
The asymmetry between the $E_Z^{x}$ and $E_Z^{y}$ terms comes from the fact that in the presence of Rashba SOC, the only remaining symmetries are TRS and
the reflection symmetry $\mathcal{R}_{K/K'}(y)=\sigma_x s_y \tau_0$, and thus, a Zeeman term proportional to $s_y$ cannot open a gap because it preserves $\mathcal{R}_{K/K'}(y)$. 
In turn, the Zeeman term proportional to $s_x$ removes the remaining symmetries (TRS and $\mathcal{R}_{K/K'}(y)$) and therefore, it is the only term responsible for the shifted crossing point.
Then, adding perturbatively $H_\text{R}^1(q_y)$, the Rashba $q_y$-dependent part given by Eq.~\eqref{rashbak1}, a gap two orders of magnitude smaller than the Zeeman energy $E_Z^x$ opens, that is, 
\begin{align}
\Delta_\text{ac}&= 2|\langle \Psi_+| H_\text{R}^1 |\Psi_-\rangle|=E_Z^x \frac{12\lambda_\text{R}^2 }{\lambda_\text{SOC}(V_{pp\pi}^1-V_{pp\sigma}^1)},\nonumber
\end{align}
which is in essence a second order process in $\lambda_\text{R}$, and explains why the gap opening becomes reduced two orders of magnitude $\Delta_\text{ac}\sim 10^{-2}E_Z^x$.

\section{Magnetoconductance}\label{Sec:magcond}
\subsection{General formalism}\label{Sec:magcond_gen}
In order to compute the magnetoconductance of the edge states inside the bulk band gap, we employ a 1D edge channel model. We first compute the retarded Green's function $\mathcal{G}^0_{nn',\alpha\beta}(k,E)$ of a nanoribbon at a fixed energy $E$, which is given by
\begin{equation}\label{Eq:GenGF}
\mathcal{G}^0_{nn',\alpha\beta}(k)=\sum\limits_{j}\frac{(\chi_{jk})_{n\alpha}(\chi^\dagger_{jk})_{n'\beta}}{E+\i0^+-\epsilon_j(k)}.
\end{equation}
Here, $k$ denotes the momentum along the nanoribbon (that is, $k=k_x$ for ZZ nanoribbons and $k=k_y$ for AC nanoribbons in the convention used in Sec.~\ref{Sec:TBnr}), $n$ and $n'$ the transverse lattice sites across the width of the nanoribbon, that is, $n,n'=1\dots,N$, and $\alpha$ and $\beta$ label the 8 combinations for states on sublattice $A/B$ with orbital $p_x/p_y$ and spin $\uparrow/\downarrow$. The eigenenergies $\epsilon_j(k)$ and eigenstates $\chi_{jk}$ of a nanoribbon are determined by
\begin{equation}\label{Eq:TBSol}
\sum\limits_{n'\beta}\left[H^\mathrm{zz/ac}_{nn'}(k)+H_\mathrm{Z}\delta_{n,n'}\right]_{\alpha\beta}(\chi_{jk})_{n'\beta}=\epsilon_j(k)(\chi_{jk})_{n\alpha},
\end{equation}
where $j$ is a subband index and the matrices $H^\mathrm{zz/ac}_{nn'}(k)+H_\mathrm{Z}\delta_{n,n'}$ are defined in Sec.~\ref{Sec:TBnr}.

Next, we restrict ourselves to one of the nanoribbon edges and introduce continuous coordinates along this edge, denoted by coordinates $x,x'$ for both ZZ or AC edges subsequently. Then, we conduct a Fourier transform of the $8N\times8N$ matrix $\tilde{\mathcal{G}}^0_{nn',\alpha\beta}(k,E)$ with respect to $k$,
\begin{equation}\label{Eq:GF_FT}
\tilde{\mathcal{G}}^0_{nn',\alpha\beta}(x-x')=\int\frac{\d k}{2\pi}\e^{\i k(x-x')}\tilde{\mathcal{G}}^0_{nn',\alpha\beta}(k).
\end{equation}
In the following, we will consider only energies inside the bulk gap, that is, a situation where there are only two counter-propagating states at the edge considered~\footnote{In total, there are four states at a given energy inside the bulk band gap. However, two of these states are located at the edge opposite to the one considered. Due to their strongly localized nature in bismuthene on SiC, states on opposite edges do not couple to each other. Hence, we can ignore the two states on the opposite edge in our model.}. Using the residue theorem, Eq.~(\ref{Eq:GF_FT}) then yields
\begin{equation}\label{Eq:GF_1D}
\begin{array}{ll}
\tilde{\mathcal{G}}^0_{nn',\alpha\beta}(x-x')=&\mathcal{G}^0_{+}(x-x')(\chi_+)_{n\alpha}(\chi^\dagger_+)_{n'\beta}\\
&+\mathcal{G}^0_{-}(x-x')(\chi_-)_{n\alpha}(\chi^\dagger_-)_{n'\beta}
\end{array}
\end{equation}
with the helical states described by
\begin{equation}\label{Eq:GF_1Ddef}
\mathcal{G}^0_{\pm}(x-x')=\frac{\e^{\i k_\pm(x-x')}\theta\left[\pm(x-x')\right]}{\i\hbar v_\pm},
\end{equation}
where the momenta $k_j=k_j(E)$ are determined by $\epsilon_j(k_j)=E$ with $j=\pm$ referring to the right-moving ($+$) and left-moving ($-$) states. The corresponding eigenstates and absolute values of the group velocities at $k=k_\pm$ are denoted as $\chi_\pm$ and $v_\pm=\left|\partial\epsilon_\pm/\partial k|_{k=k_\pm}/\hbar\right|$, respectively, with $\partial\epsilon_+/\partial k|_{k=k_+}>0$ and $\partial\epsilon_-/\partial k|_{k=k_-}<0$. Note that $k_\pm$, $v_\pm$, $\chi_\pm$ all depend on $E$ and the magnetic field $\bm{B}$.

Projecting the Green's function given by Eq.~(\ref{Eq:GF_1D}) on the propagating modes $\chi_\pm$ then yields an effective 1D Green's function for the helical edge channels. To derive the Green's function $\tilde{\mathcal{G}}_{nn',\alpha\beta}(x,x')$ in the presence of an impurity potential $\tilde{\mathcal{V}}_{nn',\alpha\beta}(x)$, we make use of the Dyson equation
\begin{widetext}
\begin{equation}\label{Eq:DyEq}
\tilde{\mathcal{G}}_{nn',\alpha\beta}(x,x')=\tilde{\mathcal{G}}^0_{nn',\alpha\beta}(x-x')+\sum\limits_{\alpha'\beta'}\sum\limits_{\tilde{n}\tilde{n}'}\int\d\tilde{x}\,\tilde{\mathcal{G}}^0_{n\tilde{n},\alpha\alpha'}(x-\tilde{x})\tilde{\mathcal{V}}_{\tilde{n}\tilde{n}',\alpha'\beta'}(\tilde{x})\tilde{\mathcal{G}}_{\tilde{n}'n',\beta'\beta}(\tilde{x},x')
\end{equation}
and expand $\tilde{\mathcal{G}}_{nn',\alpha\beta}(x,x')=\sum\limits_{i,j=\pm}\mathcal{G}_{ij}(x,x')(\chi_i)_{n\alpha}(\chi^\dagger_j)_{n'\beta}$. Introducing $\mathcal{V}_{ij}(x)=\sum\limits_{\alpha\beta}\sum\limits_{nn'}(\chi^\dagger_i)_{n\alpha}\tilde{\mathcal{V}}_{nn',\alpha\beta}(x)(\chi_j)_{n'\beta}$, we then rewrite Eq.~(\ref{Eq:DyEq}) as an integral equation with $2\times2$ matrices with respect to the two helical modes $i,j=\pm$~\footnote{Note that Eq.~(\ref{Eq:DyEqRed}) follows exactly from Eq.~(\ref{Eq:DyEq}) only if $\chi_+$ and $\chi_-$ are orthogonal to each other.},
\begin{equation}\label{Eq:DyEqRed}
\begin{array}{l}
\left(\begin{array}{cc} \mathcal{G}_{++}(x,x') & \mathcal{G}_{+-}(x,x') \\
                        \mathcal{G}_{-+}(x,x') & \mathcal{G}_{--}(x,x') \\ \end{array}\right)=\\
\left(\begin{array}{cc} \mathcal{G}^0_+(x-x') & 0 \\
                        0 & \mathcal{G}^0_-(x-x') \\ \end{array}\right)+
\int\d\tilde{x}\left(\begin{array}{cc} \mathcal{G}^0_+(x-\tilde{x}) & 0 \\
                                       0 & \mathcal{G}^0_-(x-\tilde{x}) \\ \end{array}\right)
\left(\begin{array}{cc} \mathcal{V}_{++}(\tilde{x}) & \mathcal{V}_{+-}(\tilde{x}) \\
                        \mathcal{V}_{-+}(\tilde{x}) & \mathcal{V}_{--}(\tilde{x}) \\ \end{array}\right)
\left(\begin{array}{cc} \mathcal{G}_{++}(\tilde{x},x') & \mathcal{G}_{+-}(\tilde{x},x') \\
                        \mathcal{G}_{-+}(\tilde{x},x') & \mathcal{G}_{--}(\tilde{x},x') \\ \end{array}\right).
\end{array}
\end{equation}
Instead of solving the integral equation~(\ref{Eq:DyEqRed}), it is often more convenient to solve the corresponding differential equation. Taking the derivative of Eq.~(\ref{Eq:DyEqRed}) with respect to $x$ and using $\mathcal{G}^0_{\pm}(x-x')$ as given in Eq.~(\ref{Eq:GF_1Ddef}), we obtain
\begin{equation}\label{Eq:EffDE}
\left[
\left(\begin{array}{cc} \hbar v_+\left(k_++\i\partial_x\right) & 0 \\
                        0 & -\hbar v_-\left(k_-+\i\partial_x\right) \\ \end{array}\right)-
\left(\begin{array}{cc} \mathcal{V}_{++}(x) & \mathcal{V}_{+-}(x) \\
                        \mathcal{V}_{-+}(x) & \mathcal{V}_{--}(x) \\ \end{array}\right)																
\right]\left(\begin{array}{cc} \mathcal{G}_{++}(x,x') & \mathcal{G}_{+-}(x,x') \\
                        \mathcal{G}_{-+}(x,x') & \mathcal{G}_{--}(x,x') \\ \end{array}\right)=
\left(\begin{array}{cc} 1 & 0 \\
                        0 & 1 \\ \end{array}\right)\delta(x-x').
\end{equation}
\end{widetext}
Here, $\mathcal{G}$ and $\mathcal{V}$ denote projections of the full TB Green's functions and impurity potentials $\tilde{\mathcal{G}}$ and $\tilde{\mathcal{V}}$ to the subspace spanned by the two propagating modes $j=\pm$. We model the full impurity potential $\tilde{\mathcal{V}}_{nn',\alpha\beta}(x)$ by point impurities located at the edge,
\begin{equation}\label{Eq:ImpScal}
\tilde{\mathcal{V}}_{nn',\alpha\beta}(x)=\sum\limits_{l=1}^{N_\mathrm{i}}V_l\delta_{n,n'}\delta_{n,N_\mathrm{e}}\left[\pi_0\,\frac{\overline{\sigma}_0+\overline{\sigma}_z}{2}\,s_0\right]_{\alpha\beta}\delta(x-x_l),
\end{equation}
\begin{equation}\label{Eq:ImpMag}
\tilde{\mathcal{V}}_{nn',\alpha\beta}(x)=\sum\limits_{l=1}^{N_\mathrm{i}}V_l\delta_{n,n'}\delta_{n,N_\mathrm{e}}\left[\pi_0\,\frac{\overline{\sigma}_0+\overline{\sigma}_z}{2}\,s_x\right]_{\alpha\beta}\delta(x-x_l)
\end{equation}
for scalar and magnetic impurities, respectively. Here, $N_\mathrm{i}$ denotes the number of impurities, $x_l$ the position of the $l$th impurity along the edge (with $x_1\leq...\leq x_{N_\mathrm{i}}$), $N_\mathrm{e}$ the transverse coordinate of the edge considered, $V_l$ the strength of the scalar or magnetic impurities, and $\pi_i$, $\overline{\sigma}_i$, and $s_i$ are Pauli matrices with respect to the orbitals $p_{x/y}$, the sublattice $A/B$ and spin. Hence, the elements of the $2\times2$ matrix $\underline{\underline{\mathcal{V}}}(x)$ have the form $\mathcal{V}_{ij}(x)=\sum\limits_{l}(M_l)_{ij}\delta(x-x_l)$, where the components $(M_l)_{ij}$ of the $2\times2$ matrices $\underline{\underline{M}}_l$ are computed from Eqs.~(\ref{Eq:ImpScal}) and~(\ref{Eq:ImpMag}). Here,
\begin{equation}\label{Eq:MImpScal}
(M_l)_{ij}=V_l\sum\limits_{\alpha\beta}(\chi^\dagger_i)_{N_\mathrm{e}\alpha}\left[\pi_0\,\frac{\overline{\sigma}_0+\overline{\sigma}_z}{2}\,s_0\right]_{\alpha\beta}(\chi_j)_{N_\mathrm{e}\beta}
\end{equation}
for scalar impurities and
\begin{equation}\label{Eq:MImpMag}
(M_l)_{ij}=V_l\sum\limits_{\alpha\beta}(\chi^\dagger_i)_{N_\mathrm{e}\alpha}\left[\pi_0\,\frac{\overline{\sigma}_0+\overline{\sigma}_z}{2}\,s_x\right]_{\alpha\beta}(\chi_j)_{N_\mathrm{e}\beta}
\end{equation}
for magnetic impurities.

Equation~(\ref{Eq:EffDE}) for the $2\times2$ matrix $\underline{\underline{\mathcal{G}}}(x,x')$ with components $\mathcal{G}_{ij}(x,x')$ can then be solved using the boundary conditions
\begin{equation}\label{GFbc1}
\underline{\underline{\mathcal{G}}}(x'^+,x')-\underline{\underline{\mathcal{G}}}(x'^-,x')=
\left(\begin{array}{cc} \frac{1}{\i\hbar v_+} & 0 \\
                        0 & -\frac{1}{\i\hbar v_-} \\ \end{array}\right),
\end{equation}
\begin{equation}\label{GFbc2}
\underline{\underline{\mathcal{G}}}(x_l^+,x')=\exp\left[
\left(\begin{array}{cc} \frac{1}{\i\hbar v_+} & 0 \\
                        0 & -\frac{1}{\i\hbar v_-} \\ \end{array}\right)\underline{\underline{M}}_l
\right]
\underline{\underline{\mathcal{G}}}(x_l^-,x'),
\end{equation}
\begin{equation}\label{GFbc3}
\underline{\underline{\mathcal{G}}}(x,x')=
\left(\begin{array}{cc} \e^{\i k_+(x-x_0)} & 0 \\
                        0 & \e^{\i k_-(x-x_0)} \\ \end{array}\right)
\underline{\underline{\mathcal{G}}}(x_0,x'),
\end{equation}
which have been obtained after (path-ordered) integration of Eq.~(\ref{Eq:EffDE}). Equations~(\ref{GFbc2}) and~(\ref{GFbc3}) are valid for $x'\neq x,x_0,x_l$ and describe scattering at the impurity at $x=x_l$ and propagation from $x_0$ to $x$, respectively. The transmission of a right-moving edge state through all the impurities can then be computed from the Fisher-Lee relation as
\begin{equation}\label{Transmission}
\begin{array}{ll}
T&=\left|\i\hbar v_+\mathcal{G}_{++}(x_{N_\mathrm{i}}+L,x_1-L)\right|^2\\
&=\left|\mathcal{M}_{++}-\frac{\mathcal{M}_{+-}\mathcal{M}_{-+}}{\mathcal{M}_{--}}\right|^2
\end{array}
\end{equation}
with $L>0$ \cite{Fisher1981:PRB}. Here, $\mathcal{M}_{ij}$ are components of the matrix
\begin{widetext}
\begin{equation}\label{TMatrix}
\begin{array}{ll}
\underline{\underline{\mathcal{M}}}=&\left(\begin{array}{cc} \e^{\i k_+L} & 0 \\
                        0 & \e^{\i k_-L} \\ \end{array}\right)
\prod\limits_{l=1}^{N_\mathrm{i}}\left\{
\exp\left[
\left(\begin{array}{cc} \frac{1}{\i\hbar v_+} & 0 \\
                        0 & -\frac{1}{\i\hbar v_-} \\ \end{array}\right)\underline{\underline{M}}_l
\right]
\left(\begin{array}{cc} \e^{\i k_+(x_l-x_{l-1})} & 0 \\
                        0 & \e^{\i k_-(x_l-x_{l-1})} \\ \end{array}\right)
\right\},
\end{array}
\end{equation}
\end{widetext}
if we define $x_0=x_1-L$. The corresponding edge-state conductance can then be obtained as
\begin{equation}\label{Conductance}
G(E,\bm{B})=\frac{e^2}{h}T(E,\bm{B})
\end{equation}
from the transmission $T(E,\bm{B})$.

\begin{figure}[t]
\centering
\includegraphics*[width=8.5cm]{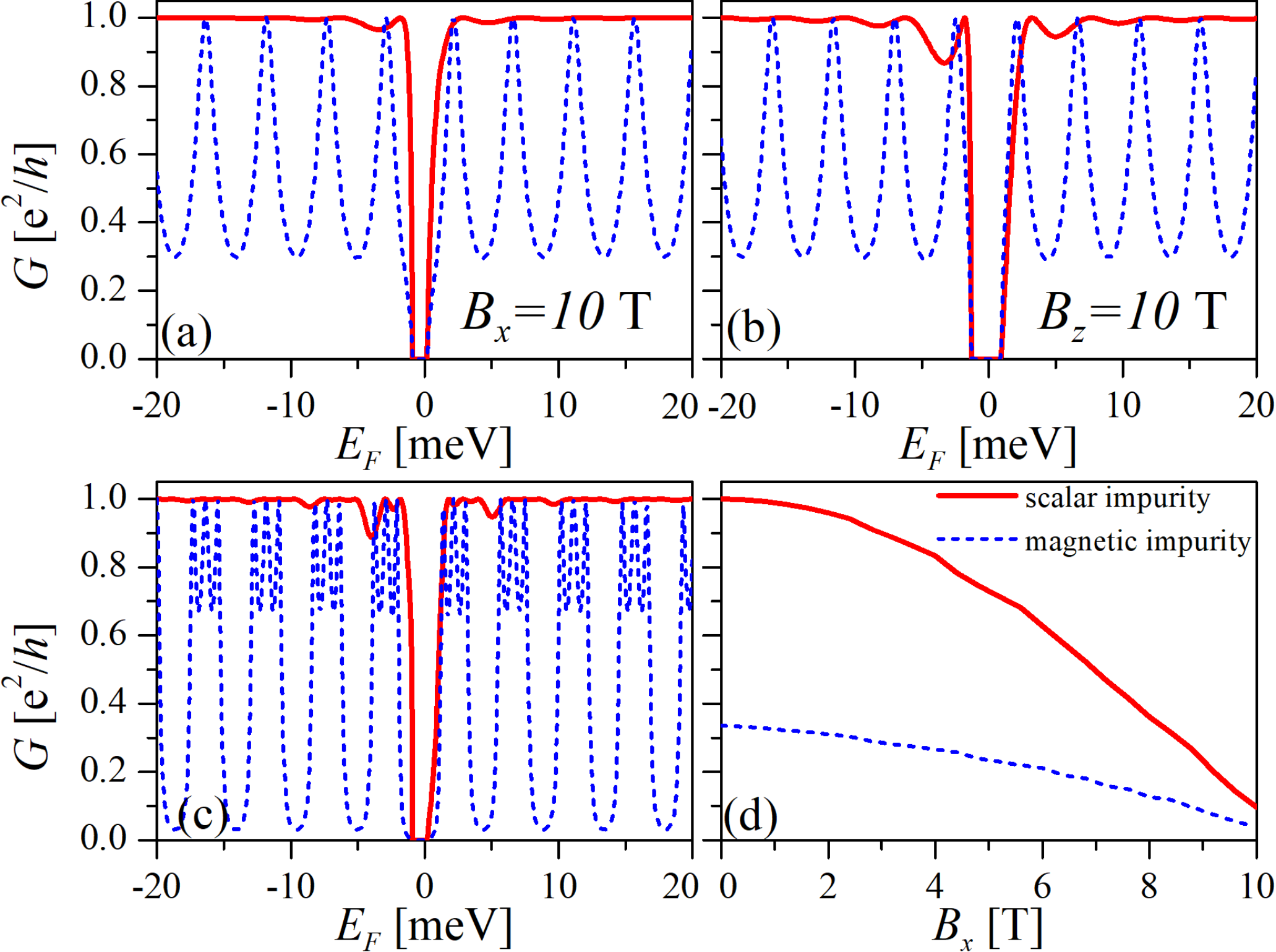}
\caption{Conductance $G$ of a single ZZ edge as a function of $E_F$ for (a) $B_x=10$ T and (b) $B_z=10\,$T with two point-like scalar or magnetic impurities present at the edge. (c) Same as~(a), but with four impurities. (d) Dependence of $G$ on $B_x$ for $E_F=250\,\mu$eV and two impurities. In all panels, the impurities are separated by $d_\mathrm{imp}=100\,$nm from each other and have a strength of $V_\mathrm{imp}=10\,${eVnm} each.}\label{fig:ConductanceZZ}
\end{figure}

Figure~5 in the main text shows the magnetoconductance for a single AC edge in the presence of impurities, where each scalar and magnetic impurity has the same strength $V_l=V_\mathrm{imp}$ in Eqs.~(\ref{Eq:ImpScal})-(\ref{Eq:MImpMag}) and impurities are separated from each other by a distance $d_\mathrm{imp}$. The results for a single ZZ edge are qualitatively similar, although the gap $\Delta_\text{zz}$ opened by an in-plane field is now much larger and of the order of meV, as illustrated in Fig.~\ref{fig:ConductanceZZ} in this Supplementary.

\subsection{Perfectly spin-polarized states}\label{Sec:magcond_spinpolarized}

While Fig.~5 in the main text and Fig.~\ref{fig:ConductanceZZ} in this Supplementary have been calculated numerically with $\lambda_\text{R}\neq0$, relatively simple analytical expressions can be derived if the two counter-propagating modes $\chi_{\pm}$ are perfectly spin-polarized along the $z$-direction and orthogonal to each other, that is, for $\lambda_\text{R}=0$. Then,
\begin{equation}\label{Eq:ImpScalSimple}
\underline{\underline{M}}_l=\left(\begin{array}{cc} (M_l)_{++} & 0 \\
                             0 & (M_l)_{++} \\ \end{array}\right)
\end{equation}
for a scalar impurity and
\begin{equation}\label{Eq:ImpMagSimple}
\underline{\underline{M}}_l=\left(\begin{array}{cc} 0 & (M_l)_{+-} \\
                             (M_l)_{+-} & 0 \\ \end{array}\right)
\end{equation}
for a magnetic impurity with magnetic moment in $x$-direction. In Eq.~(\ref{Eq:ImpScalSimple}), the off-diagonal elements $(M_l)_{+-}$ and $(M_l)_{-+}$ as computed from Eq.~(\ref{Eq:MImpScal}) vanish for perfectly $z$-spin-polarized edge states, while $(M_l)_{++}=(M_l)_{--}$ for the diagonal elements. For magnetic impurities, on the other hand, Eq.~(\ref{Eq:MImpMag}) yields $(M_l)_{++}=(M_l)_{--}=0$ and finite $(M_l)_{+-}=(M_l)_{-+}$, and we then obtain Eq.~(\ref{Eq:ImpMagSimple}).

In the case of scalar impurities, Eq.~(\ref{TMatrix}) together with Eq.~(\ref{Eq:ImpScalSimple}) yields a diagonal $\underline{\underline{\mathcal{M}}}$, where scattering at an impurity at $x=x_l$ only adds a phase $\mp\i(M_l)_{++}/(\hbar v_\pm)$ for right-/left-movers and consequently $T=|\mathcal{M}_{++}|^2=1$. On the other hand, in the presence of magnetic impurities $\underline{\underline{\mathcal{M}}}$ is computed from Eqs.~(\ref{TMatrix}) and~(\ref{Eq:ImpMagSimple}) and contains off-diagonal terms due to back-scattering at the impurities that reduce $T$.

For example, the edge channel transmissions in the presence of one and two magnetic impurities are given by
\begin{equation}\label{Eq:Trans1MI}
T=\frac{1}{\cosh^2Z_1}
\end{equation}
and
\begin{equation}\label{Eq:Trans2MI}
T=\left|\frac{\e^{\i k_+d_\mathrm{imp}}}{\cosh Z_1\cosh Z_2+\e^{\i(k_+-k_-)d_\mathrm{imp}}\sinh Z_1\sinh Z_2}\right|^2,
\end{equation}
respectively. Here, the distance between the two impurities is denoted as $d_\mathrm{imp}$ and we have introduced $Z_l=(M_l)_{+-}/(\hbar\sqrt{v_+v_-})$ for the two impurities labeled by $l=1,2$. Equation~(\ref{Eq:Trans2MI}) is equivalent to the transmission through a quantum-dot structure based on QSH edge states and two magnetic tunneling barriers~\cite{Timm2012:PRB}.

If $Z_l\gg1$, Eqs.~(\ref{Eq:Trans1MI}) and~(\ref{Eq:Trans2MI}) scale as $T\propto\e^{-2Z_1}$ and $T\propto\e^{-2Z_1}\e^{-2Z_2}$, respectively. For an arbitrary number of impurities $N_\mathrm{i}$ with $Z_l\gg1$, this can be extended to
\begin{equation}\label{Eq:TransNiMIlargeZ}
T\propto\prod\limits_{l=1}^{N_\mathrm{i}}\e^{-2Z_l}.
\end{equation}
In this limit, Eq.~(\ref{Eq:TransNiMIlargeZ}) implies that the effect of each magnetic impurity $l$ is to reduce the edge-state transmission $T$ (and consequently $G$) by $\e^{-2Z_l}$. If the impurity strength is the same for each impurity, $Z_l\equiv Z$, Eq.~(\ref{Eq:TransNiMIlargeZ}) reduces to $T\propto\e^{-2ZN_\mathrm{i}}$ and $T$ decays exponentially with the number of impurities $N_\mathrm{i}$. Although Eq.~(\ref{Eq:TransNiMIlargeZ}) has been derived only for perfectly spin-polarized edge states and $Z_l\gg1$, it also provides a good estimate for $T$ (and consequently $G$) in the presence of Rashba SOC and magnetic fields.

\section{Computation of the Magnetic Moment Tensors}\label{Sec:gfactor}

\begin{figure}[t]
\centering
\includegraphics*[width=8.5cm]{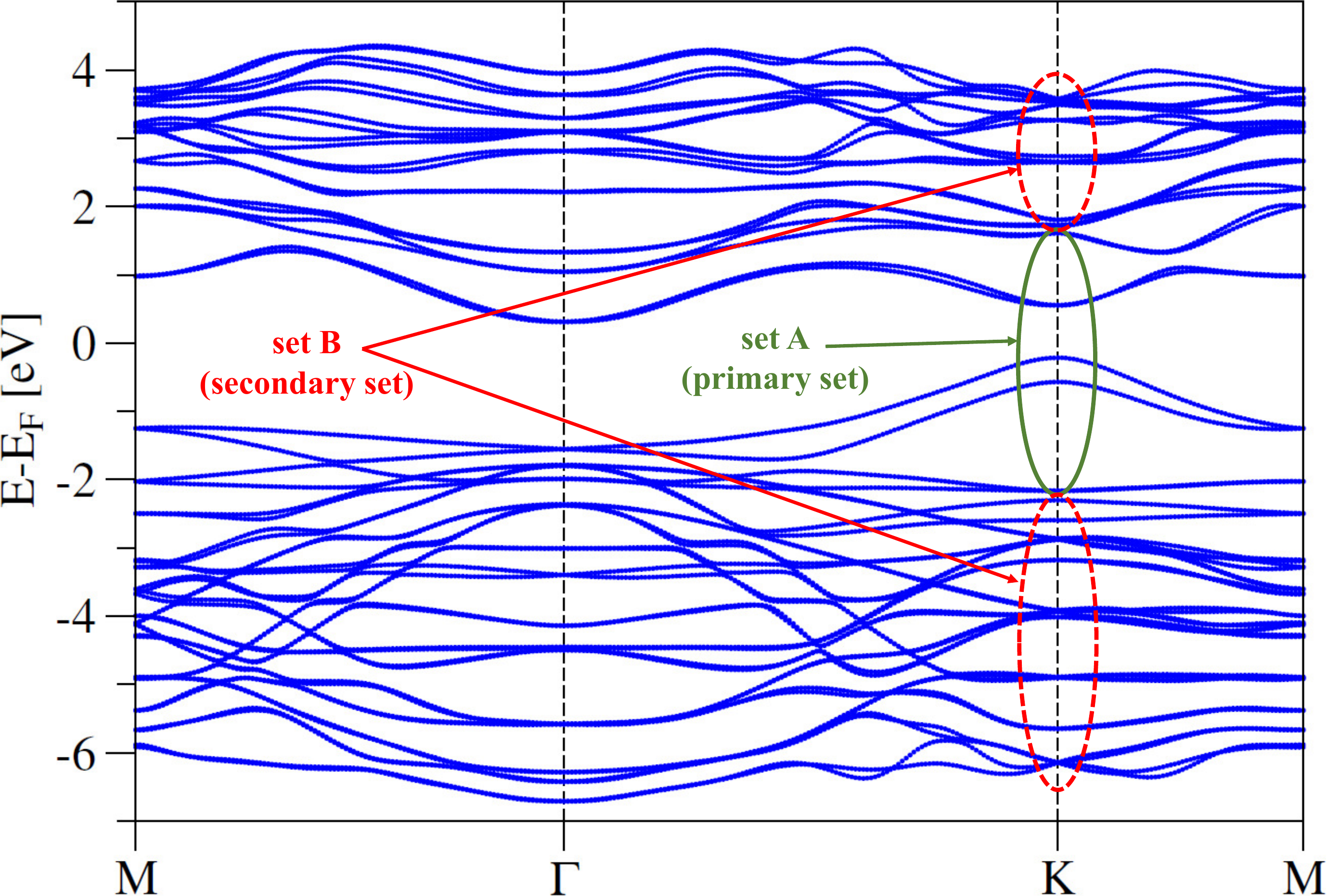}
\caption{Band structure of bismuthene on SiC as calculated using an ab-initio-based tight-binding model.Here, the primary set A as well as the secondary set B have been indicated.}\label{fig:TBDFT}
\end{figure}

We calculate the effective magnetic moment tensor $\tilde{\bm{\mu}}$ of the low-energy valence and conduction bands (main set, called set A in Fig.~\ref{fig:TBDFT}) and its renormalization due to the presence of higher energy bands (secondary set, called set B in Fig.~\ref{fig:TBDFT}). To this aim, we employ second-order quasi-degenerate perturbation theory (L\"{o}wdin perturbation theory), from which we obtain $\tilde{\bm{\mu}}$ at the $\bm{K}$ point~\cite{Winkler2003,Graf1995:PRB},
\begin{equation}\label{gfactor}
\begin{array}{l}
\left(\tilde{\mu}_\alpha\right)_{jj'}=\\
\frac{\mu_\mathsmaller{\mathrm{B}}}{2}\Bigg[g_0\left(s_\alpha\right)_{jj'}\\
-\i m_0\sum\limits_l\left(\frac{\sum\limits_{\beta\gamma}\epsilon_{\alpha\beta\gamma}\left(v_\beta\right)_{jl}\left(v_\gamma\right)_{lj'}}{\epsilon_j(\bm{K})-\epsilon_l(\bm{K})}+\frac{\sum\limits_{\beta\gamma}\epsilon_{\alpha\beta\gamma}\left(v_\beta\right)_{jl}\left(v_\gamma\right)_{lj'}}{\epsilon_{j'}(\bm{K})-\epsilon_l(\bm{K})}\right)\Bigg]
\end{array}
\end{equation}
with the Bohr magneton $\mu_\mathsmaller{\mathrm{B}}$, the bare electron mass $m_0$, and the bare $g$ factor $g_0$. Here, $\alpha$, $\beta$ and $\gamma$ each label the directions $x$, $y$ and $z$, $\epsilon_{\alpha\beta\gamma}$ is the Levi-Civita tensor, $j$, $j'$ denote the different quasi-degenerate bands as the main set considered, while $l$ denotes the bands outside this quasi-degenerate set.

Equation~(\ref{gfactor}) arises from the coupling between the primary set A (consisting of the 8 bands of our TB model) and the secondary set B by a magnetic field $\bm{B}$. The renormalized Zeeman term~(\ref{gfactor}) is then an $8\times8$ matrix that acts only on the subspace of set A, but that also describes the effect of set B on set A. Since in our case there are additional bands close in energy to the 8 bands of our TB model/primary set (see Fig.~\ref{fig:TBDFT}), it is important to take the influence of these additional bands into account via Eq.~(\ref{gfactor}). This procedure, however, describes only the coupling between sets A and B by a magnetic field $\bm{B}$, but not the orbital effects of $\bm{B}$ within set A. In order to describe these orbital effects, we have to include the Peierls phase in our 8-band TB model.

We use an ab-initio-based TB Hamiltonian to reproduce the energies $\epsilon_{j/l}(\bm{K})$ and the spin and velocity matrix elements, $(s_\alpha)_{jj'}$ and $(v_\beta)_{jl}$, at the $\bm{K}$ point obtained by density functional theory (DFT)~\cite{Hohenberg1964:PR,Kohn1965:PR}. More specifically, we reproduce 20 conduction bands and 32 valence bands from the total spectrum (see Fig.~\ref{fig:TBDFT}), employing the Vienna Ab initio simulation package (VASP)~\cite{Kresse1996:PRB} with the projector augmented-wave pseudopotential~\cite{Bloechl1994:PRB}. In addition, we treat the exchange-correlation energy within the generalized gradient approximation of Perdew, Burke and Ernzerhof (PBE)~\cite{Perdew1996:PRL}. In our calculations, the SOC of the electrons was considered self-consistently and the lattice constant of SiC was taken as 5.35 \AA. The Wannierization was carried out with a $\Gamma$-centered Monkhorst-Pack special $k$-point method~\cite{Monkhorst1976:PRB} with a $9\times9\times1$ grid using the \textsc{wannier90} package~\cite{Marzari1997:PRB}. The supercell contains 2 Bi atoms and 6 Si atoms, and Bi-centered $s$ and $p$ orbitals and Si-centered $p$ orbitals, each for spin $\uparrow$ and $\downarrow$, are chosen as the basis of the $52\times52$ TB Hamiltonian.

Turning back to Eq.~\eqref{gfactor}, we divide the 52 energy bands into two different sets: the main set A [$j,j'$ in Eq.~(\ref{gfactor})] is composed of the 8 low-energy bands (4 lowest conduction bands and the 4 highest valence bands), and the remaining 44 bands comprise the secondary set B [$l$ in Eq.~(\ref{gfactor})]. Then, by means of Eq.~\eqref{gfactor} we obtain an $8\times8$ matrix $\tilde{\mu}_\alpha$ ($\alpha=x,y,z$) with respect to the main set basis. Next, we identify this basis with the eigenstates at $\bm{K}$ of the $8\times8$ TB model given by Eqs.~(1)-(3) in the main text [Eqs.~(\ref{Eq:TotalHamiltonian})-(\ref{Eq:HamiltonianRashbaParameters})]~\footnote{Since the $8\times8$ TB model given by Eqs.~(1)-(3) in the main text contains only $p_x$ and $p_y$ orbitals from the Bi atoms, this identification is only an approximate one. From the full $52\times52$ ab-initio-based TB model, we find that the contribution of the Bi $p_x$ and $p_y$ orbitals amounts to around $80\%$ for the two low-energy conduction and valence bands at $\bm{K}$, that is, the bands from which the QSH edge states originate, with the remaining $20\%$ due to Bi $s$ and $p_z$ orbitals and Si $p$ orbitals.}. Hence, in order to obtain $\mu_\alpha$, the magnetic moment tensor in the basis of Eqs.~(1)-(3), we transform the $8\times8$ matrix computed with Eq.~(\ref{gfactor}) and the DFT-based TB model by $\mu_\alpha=U\,\tilde{\mu}_\alpha\,U^t$, where the unitary matrix $U$ describes the transformation that diagonalizes the $8\times8$ TB model given by Eqs.~(1)-(3) at $\bm{K}$.

For an in-plane magnetic field, the orbital corrections in Eq.~(\ref{gfactor}) are very small [contributions from $\left(v_z\right)_{jl}$ are several orders of magnitude smaller than the ones from $\left(v_{x/y}\right)_{jl}$]. As a consequence, we do not find a significant renormalization of the bare Zeeman term for in-plane $\bm{B}$ and obtain the tensors
\begin{equation}\label{eq:gxtensor}
\mu_{x/y}=\frac{g_0\mu_\mathsmaller{\mathrm{B}}}{2}\bm{1}\otimes s_{x/y},
\end{equation}
in the basis $\ket{p^A_{x\uparrow}}$, $\ket{p^A_{y\uparrow}}$, $\ket{p^B_{x\uparrow}}$, $\ket{p^B_{y\uparrow}}$, $\ket{p^A_{x\downarrow}}$, $\ket{p^A_{y\downarrow}}$, $\ket{p^B_{x\downarrow}}$, $\ket{p^B_{y\downarrow}}$, where $s_\alpha$ are Pauli spin matrices and $\bm{0}$ and $\bm{1}$ are the $4\times4$ zero and identity matrices, respectively.

For an out-of-plane field, on the other hand, the orbital corrections are more pronounced and we find significant corrections to the bare magnetic moment. After the transformation to the basis order $\ket{p^A_{x\uparrow}}$, $\ket{p^A_{y\uparrow}}$, $\ket{p^B_{x\uparrow}}$, $\ket{p^B_{y\uparrow}}$, $\ket{p^A_{x\downarrow}}$, $\ket{p^A_{y\downarrow}}$, $\ket{p^B_{x\downarrow}}$, $\ket{p^B_{y\downarrow}}$, we find
\begin{widetext}
\begin{equation}\label{eq:gztensor}
\footnotesize
\mu_z=\frac{\mu_\mathsmaller{\mathrm{B}}}{2}\left(\begin{array}{cccccccc}
 0.44 & -0.56+2.16\i & 1.55-0.38\i & 0.60-1.26\i & 0.68+1.91\i & 1.47+1.61\i & -0.91-1.85\i & 2.69+2.68\i\\
 -0.56-2.16\i & 2.15 & 0.16+1.10\i & 1.50-0.23\i & -1.59+0.16\i & -2.60+1.07\i & 1.64+0.69\i & -3.97+1.95\i\\
 1.55+0.38\i & 0.16-1.10\i & -0.37 & 1.30+0.84\i & 3.76-1.77\i & -0.98-4.09\i & -0.18+1.87\i & -2.97-2.99\i\\
 0.60+1.26\i & 1.50+ 0.23\i & 1.30-0.84\i & 4.40 & -1.01-1.90\i & -3.23+0.91\i & 1.95+0.80\i & -2.80+3.09\i\\
 0.68-1.91\i & -1.59-0.16\i & 3.76+1.77\i & -1.01+1.90\i & -3.68 & 0.58+0.81\i & 1.01+0.11\i & 0.36+0.73\i\\
 1.47-1.61\i & -2.59-1.07\i & -0.98+4.09\i & -3.23-0.91\i & 0.58-0.81\i & -0.82 & -2.38-0.26\i & -0.06-1.28\i\\
 -0.91+1.85\i & 1.64-0.69\i & -0.18-1.87\i & 1.95-0.80\i & 1.01-0.11\i & -2.38+0.26\i & -1.52 & -0.49+0.96\i\\
 2.69-2.68\i & -3.97-1.95\i & -2.97+2.99\i & -2.80-3.09\i & 0.36-0.73\i & -0.06+1.28\i & -0.49-0.96\i & -2.71\\
 \end{array}\right).
\end{equation}
\end{widetext}
Note that, although Eq.~(\ref{eq:gztensor}) describes the response to an out-of-plane $\bm{B}$, $\mu_z$ contains terms in the off-diagonal spin blocks. Due to strong SOC~\footnote{If no SOC was taken into account, Eq.~(\ref{eq:gztensor}) would not contain any terms mixing spin.}, most bands in sets A and B are not quantized along the $z$-direction, but along different directions for different bands. Coupling of such bands with different spin orientations then yields the off-diagonal terms in Eq.~(\ref{eq:gztensor}). Then, the Zeeman term in the basis of the $8\times8$ TB model given by Eqs.~(1)-(3) in the main text [Eqs.~(\ref{Eq:TotalHamiltonian})-(\ref{Eq:HamiltonianRashbaParameters})] reads
\begin{equation}
H_\mathrm{Z}=\bm{\mu}\cdot\bm{B}=\mu_xB_x+\mu_yB_y+\mu_zB_z,
\end{equation}
where $\mu_\alpha$ is given by Eqs.~(\ref{eq:gxtensor})-(\ref{eq:gztensor}).

\end{document}